\documentclass[aps,12pt,
    prd,
    reprint,
    onecolumn,
    superscriptaddress,
    nolongbibliography,
    nofootinbib,
    floatfix,
    notitlepage
]{revtex4-2}
\usepackage{amsmath,amssymb,amsfonts,amsthm}

\usepackage[dvipsnames]{xcolor}

\usepackage{hyperref}
\hypersetup{colorlinks=true,citecolor=Cyan,urlcolor=Red}

\usepackage{mathrsfs}
\usepackage{bbm}
\usepackage{slashed}
\DeclareSymbolFont{rsfs}{U}{rsfs}{m}{n}
\DeclareSymbolFontAlphabet{\mathrsfs}{rsfs}
\usepackage[mathscr]{eucal}
\usepackage[normalem]{ulem}
\usepackage{verbatim}
\usepackage{bm}

\usepackage{graphicx}

\newcommand{\be}{\begin{equation}}
\newcommand{\ee}{\end{equation}}
\newcommand{\bi}{\begin{itemize}}
\newcommand{\ei}{\end{itemize}}
\newcommand{\bea}{\begin{eqnarray}}
\newcommand{\eea}{\end{eqnarray}}
\newcommand{\ud}{\mathrm{d}}            

\newcommand{\LCm}{{\scriptscriptstyle -}}
\newcommand{\LCp}{{\scriptscriptstyle +}}
\newcommand{\LCpm}{{\scriptscriptstyle \pm}}

\newcommand{\LCperp}{{\scriptscriptstyle \perp}}

\usepackage{microtype}
\usepackage{siunitx}
\usepackage{centernot}
\usepackage[capitalise]{cleveref}
\crefname{equation}{}{}
\crefname{section}{Sec.}{Secs.}

\newcommand{\CA}{\ensuremath{\mathcal{A}}}
\newcommand{\CE}{\ensuremath{\mathcal{E}}}
\newcommand{\CB}{\ensuremath{\mathcal{B}}}

\DeclareMathOperator{\E}{E}

\usepackage[caption=false,labelformat=simple,listofformat=subsimple]{subfig}

\newcommand{\LL}[1]{\ensuremath{\text{LL}_{#1}}}

\begin{document}
\title{
    Reduction of order, resummation and radiation reaction
}

\author{Robin Ekman}
\email{robin.ekman@plymouth.ac.uk}
\author{Tom Heinzl}
\email{theinzl@plymouth.ac.uk}
\author{Anton Ilderton}
\email{anton.ilderton@plymouth.ac.uk}
\affiliation{Centre for Mathematical Sciences, University of Plymouth, Plymouth, PL4 8AA, UK}
\begin{abstract}
The Landau-Lifshitz equation is the first in an infinite series of approximations to the Lorentz-Abraham-Dirac equation obtained from `reduction of order'.
We show that this series is divergent, predicting wildly different dynamics at successive perturbative orders.
Iterating reduction of order \emph{ad infinitum} in a constant crossed field, we obtain an equation of motion which is free of the erratic behaviour of perturbation theory.
We show that Borel-Pad\'e resummation of the divergent series accurately reproduces the dynamics of this equation, using as little as two perturbative coefficients.
Comparing with the Lorentz-Abraham-Dirac equation, our results show that for large times the optimal order of truncation typically amounts to using the Landau-Lifshitz equation, but that this fails to capture the resummed dynamics over short times.

\end{abstract}
\maketitle

\section{Introduction}

Radiation reaction (RR) in electrodynamics becomes relevant in the presence of strong fields, where RR forces can become comparable to, or dominate, the Lorentz force.
In the presence of strong gravitational fields RR was observed decades ago in studies of the Hulse-Taylor binary pulsar~\cite{Taylor:1979zz}.
The recent observation of gravitational waves and their analysis has further highlighted the importance of RR e.g.\ at third post-Minkowskian order~\cite{Damour:2020tta}, triggering a number of investigations to clarify the subtleties involved~\cite{DiVecchia:2021bdo,Herrmann:2021tct,Bjerrum-Bohr:2021vuf}.
The direct connection between RR in gravity and in Yang-Mills theory has also recently been explored in the context of double copy~\cite{Goldberger:2016iau,delaCruz:2020bbn,Adamo:2020qru}.
Here we consider RR in classical electrodynamics~\cite{abraham1905,lorentz1909,dirac1938classical}, where it was first formulated theoretically.
Our interest is motivated by emerging experimental access to previously uncharted strong-field regimes, provided by intense lasers~\cite{Cole:2017zca,Poder:2018ifi,Wistisen:2017pgr,Nielsen:2020ufz,danson2019petawatt,Abramowicz:2021zja,Meuren:2020nbw}.

The classical equation of motion supposed to describe RR is the third-order Lorentz-Abraham-Dirac (LAD) equation ~\cite{abraham1905,lorentz1909,dirac1938classical}.
Its third-order character is in conflict with Newton-Laplace determinism, as more than two initial conditions are needed to uniquely determine a solution.
Adding an initial condition for acceleration makes the initial-value problem well posed, but leads to unphysical runaway solutions at temporal infinity.
Imposing instead  Dirac's condition of vanishing final acceleration~\cite{dirac1938classical}, one has an initial-boundary value problem, and solutions (the existence and uniqueness of which is not guaranteed~\cite{Bopp:1943,Haag:1955,Carati:1995,Carati:2021pqd}) exhibit pre-acceleration before the external field is encountered (albeit on a small time scale of about 2 fm/$c$).
Finally, both analytical and numerical solutions of LAD are hampered by the strong nonlinearities present in the fully relativistic case \cite{Plass:1961zz}.

In view of these difficulties, it is common to adopt the method of `reduction of order'.
When applied to LAD, this yields the Landau-Lifshitz (LL) equation~\cite{LandauLifshitzII}, which, being second-order in time derivatives, yields a well-posed initial-value problem and hence is free of runaway and pre-accelerating solutions~\cite{rohrlich2007classical}.
Reduction of order from LAD to LL is an example of singular perturbation theory~\cite{Bender:1999}.
It is known that the ensuing perturbative approximations to the full equation will typically miss nonlinear phenomena such as layers and bifurcation branches unless one introduces an appropriate amount of parameter fine-tuning~\cite{Bopp:1943,Haag:1955,Carati:1995,Carati:2021pqd,Heinzl:2016kzb}.
Furthermore, LL represents only the first of an infinite series of approximations to LAD obtained by \emph{iteration} of reduction of order.

With this in mind, we show here that the perturbation expansion generated by iterating reduction of order has zero radius of convergence and is asymptotic, predicting wildly different physics at successive orders of perturbation theory.
We are therefore prompted to investigate resummation of the series, as has recently been highlighted in investigations of both classical~\cite{Heinzl:2021mji} and quantum~\cite{Torgrimsson:2021wcj} radiation reaction, and in strong field problems more generally~\cite{Karbstein:2019wmj,Mironov:2020gbi,Edwards:2020npu,Dunne:2021acr}.

We focus mainly on the case of a constant crossed field (CCF) background, which is relevant to the Ritus-Narozhny conjecture on the breakdown of perturbation theory in strong fields~\cite{Fedotov:2016afw}.
We will show that the simplifications associated with the CCF case allow for an almost complete resummation of the divergent perturbative series: we are able to iterate reduction of order an \emph{infinite} number of times, casting the result as a nonlinear system of first-order, ordinary differential equations (ODEs).
While the perturbative expansion of this system is divergent, we are able to prove that the strong-field expansion is convergent, a situation reminiscent of the analogous expansions of the Heisenberg-Euler Lagrangian~\cite{Dunne:2004nc}.

Furthermore, the system can be solved numerically to high accuracy, such that the solution may be viewed as numerically `exact'.
We compare this exact result to resummed Pad\'e-Borel approximants of the perturbative expansion, showing that these can match the numerical answer to high precision using only a few terms.
We find that infinite reduction of order gives results which are free of the erratic behaviour seen in perturbation theory, and which match LAD, but without the non-perturbative, and unphysical runaways or pre-acceleration behaviour.
All our results indicate that radiation reaction rapidly drives particle motion to a regime where LL is valid; the optimal order of truncation for the divergent reduction of order expansion is therefore one.

We work in the classical theory throughout; quantum corrections are typically expected~\cite{klepikov1985radiation,bulanov2011lorentz,Blackburn:2019rfv} when high field strengths cause hard acceleration gradients and significant radiation reaction, but the inclusion of such corrections goes beyond the scope of this paper.
Our focus will therefore be on resummation and physical properties of radiation reaction equations, rather than on phenomenological predictions.

This paper is organised as follows.
In \cref{sec:LLn} we introduce reduction of order, and its iterations, of the relativistic LAD equation in a constant crossed field of arbitrary strength.
We show that the resulting series is divergent, highlight how this manifests in the physics of particle motion, and how this can be cured using Borel-Pad\'e resummation.
In \cref{sec:LL-to-all-orders}  we iterate reduction of order to all orders, obtaining an equation of motion that we dub $\LL\infty$.
We show that the strong-field expansion of $\LL\infty$ is convergent and use this to investigate the physics of the strong field regime.
In \cref{sec:stitched-stuff} we use the convergent strong-field behaviour to improve our perturbative resummation, and obtain an analytical expression for $\LL\infty$.
We then compare this with the numerical solution of LAD.
We conclude in~\cref{sec:conclusion}, suggesting a physical explanation of the divergence of perturbation theory, relating our results to the previous literature, and discussing extensions.
Throughout we use units where $c = \varepsilon_0 = \hbar = 1$.

\section{Iteration of reduction of order}
\label{sec:LLn}

Consider the orbit $x^\mu(\tau)$ of a particle, charge $e$ and mass $m$, in an external field $F_{\mu\nu}(x)$.
The orbit is parameterised by proper time $\tau$, the particle velocity is written $u^\mu = \dot{x}^\nu$, and overdots denote derivatives with respect to $\tau$.
The LAD equation of motion is, writing $f_{\mu\nu} = eF_{\mu\nu}/m$,
\begin{align}
    \label{eq:LAD}
    \dot{u}^\mu = f^{\mu\nu} u_\nu + \tau_0 \mathcal{P}^{\mu\nu} \ddot{u}_\nu \;,
\end{align}
in which $\mathcal{P}^{\mu\nu} = \eta^{\mu\nu} - u^\mu u^\nu$ projects orthogonally to $u^\mu$
and $\tau_0 = e^2/6 \pi m = 2 \alpha/3m$, $\alpha$ being the fine-structure constant, is the characteristic time scale of RR; for an electron $\tau_0 \approx \SI{6.3e-24}{s}$.
Note that~\cref{eq:LAD} is a third-order ODE for the orbit $x^\mu$; the associated difficulties of either runaways or pre-acceleration motivate the adoption of reduction of order in derivatives.
This refers to formally differentiating the LAD equation, then using the equation itself to eliminate $\ddot{u}^\mu$ (and higher derivatives) in favour of new $\tau_0$-dependent terms.
This results, in principle, in an infinite series of terms which can be truncated at any chosen order in $\tau_0$ due to the smallness of that parameter compared to relevant timescales.
Truncating at order $\tau_0$ yields the Landau-Lifshitz equation~\cite{LandauLifshitzII,Spohn:1999uf} (\LL{1} from here, for reasons which will become clear),
\begin{equation}
    \label{eq:LL}
    \dot{u}^\mu = f^{\mu\nu} u_\nu + \tau_0 f^{\mu\nu,\rho} u_\nu u_\rho + \tau_0 \mathcal{P}^{\mu\nu} f^2_{\nu\rho} u^\rho
    + \mathcal{O}(\tau_0^2)
    .
\end{equation}
Unlike~\cref{eq:LAD}, this is now a \emph{second} order ODE for $x^\mu$, hence the term 'reduction of order'.
This process can be iterated, truncating at \emph{higher orders in $\tau_0$}, but rapidly becomes complicated due to the appearance of new tensor structures and higher derivatives of the field tensor.

As we are interested in large orders in $\tau_0$ we simplify matters by limiting the number of possible terms which can appear.
To this end we begin by restricting our considerations to constant backgrounds for which all field derivatives are  identically zero, and we need make no approximation on them.
Note that, for general fields, this may be viewed as the leading order in a derivative expansion, and there are good arguments for dropping derivative terms:
these scale with powers of $\omega \tau_0$ where $\omega$ is a typical field frequency scale.
For realistic fields $\omega \tau_0 \ll 1$, and indeed it can be seen in e.g.~the exact solution of \LL{1} in a plane wave~\cite{Heintzmann_1972,PiazzaExact,DiPiazza:2021nsx} that the derivative terms only ever yield sub-leading effects in the field strength.
Iterating reduction of order to $\mathcal{O}(\tau_0^2)$~\cite{DiPiazza:2018luu,Ekman:2021vwg}, it can again be shown explicitly that derivative terms have negligible impact on the physics~\cite{Ekman:2021vwg}.
This is also consistent with the standard effective field theory ordering of field `operators' according to their dimension.
This argument comes with the caveat that the derivative terms can contribute to the \emph{non-perturbative} features of LAD~\cite{dirac1938classical,Plass:1961zz,Zhang:2013ria}, which are also of interest.

The tensor structures appearing are further simplified by taking the background to be a CCF (the zero frequency limit of a plane wave), as discussed in the introduction.
The field strength is $f_{\mu\nu} = m \CE (n_\mu \epsilon_\nu- \epsilon_\mu n_\nu)$ with $n_\mu$ lightlike, $\epsilon_\mu$ spacelike, $\epsilon\cdot n=0$, and dimensionless amplitude $\CE$.
Consider now iterating reduction of order with this background.
Because $f^3_{\mu\nu}\equiv 0$, there are only two tensor strucures, proportional to $f$ or $f^2$, which can ever appear.
For a CCF, all field invariants vanish, and the only nontrivial, dimensionless, invariant which can be constructed from the particle velocity $u_\mu$ and the field is
\begin{equation}
    \chi = \sqrt{\frac{u^\mu f^2_{\mu\nu} u^\nu}{m^2}}
     = \CE\, n \cdot u \;,
\end{equation}
which can be interpreted as the field magnitude `seen' by the particle in its instantaneous rest frame, in units of the Sauter-Schwinger field $m^2/e$~ \cite{Sauter:1931zz,Schwinger:1951nm}.
Note that $\chi$ is a composite parameter, essentially the product of field strength and particle energy.

It follows that reduction of order, iterated to \emph{arbitrary} order in $\tau_0$, will yield
an equation of the form
\begin{equation}
    \label{GENSOL}
    {\dot u}^\mu = \CA(\chi) f^{\mu\nu}u_\nu + \tau_0 \CB(\chi) (\mathcal{P} f^2)^{\mu\nu} u_\nu \;, \\
\end{equation}
with the functions $\CA$ and $\CB$ depending on $\chi$.
It is easily seen that \LL{1} corresponds to $\CA=\CB=1$ which sets the initial condition for the iteration procedure.
The equation of motion~\cref{GENSOL} is our first result.
Iterated reduction of order in more general fields may be explored by applying resummation to a perturbative expansion of the integro-differential formulation of LAD, see Ref.~\cite{klepikov1985radiation,Kazinski:2010ce}

We now turn to the explicit construction of the functions $\CA$ and $\CB$.
From here on we set the electron mass $m$ to $1$ in the text, in order to simplify our equations.
We re-instate $m$ in some figures so that the reader can easily see the physical scales.

\subsection{Perturbation theory and divergence}

Iterating reduction of order, i.e.~retaining terms of order up to and including $\tau_0^{k}$, yields a sequence of equations of form~\cref{GENSOL} which we refer to as $\LL{k}$, $k\geq 1$.
The functions $\CA$ and $\CB$ then have series expansions in $\tau_0^2 \chi^2$, given by
\begin{align}
    \label{eq:LLn}
    \CA \to \CA^{(k)} \equiv \sum_{\ell=0}^{\lfloor k/2 \rfloor}
          A_\ell (\tau_0\chi)^{2 \ell} \;,
          \qquad
    \CB \to \CB^{(k)} \equiv \sum_{\ell=0}^{\lfloor \frac{k-1}{2} \rfloor}
        B_\ell (\tau_0\chi)^{2 \ell} \;,
\end{align}
in which $A_0 = B_0 = 1$, recovering the Lorentz force and $\LL{1}$ equations, respectively.
After some algebra, reduction of order implies the following recursion relations for the coefficients $A_k$ and $B_k$:
\begin{align}
    \label{eq:AkBk}
    \begin{split}
        A_{k+1} & = -2 \sum_{\ell = 0}^k (\ell+1) A_\ell B_{k - \ell} \;,
        \qquad
        B_{k+1}  = \sum_{\ell = 0}^k A_\ell A_{k - \ell} - 2 \sum_{\ell = 0}^{k-1} (\ell+1) B_{\ell} B_{k-\ell-1}
        \,
        .
    \end{split}
\end{align}
\begin{figure}[tb]
    \centering
    \includegraphics[width=0.5\linewidth]{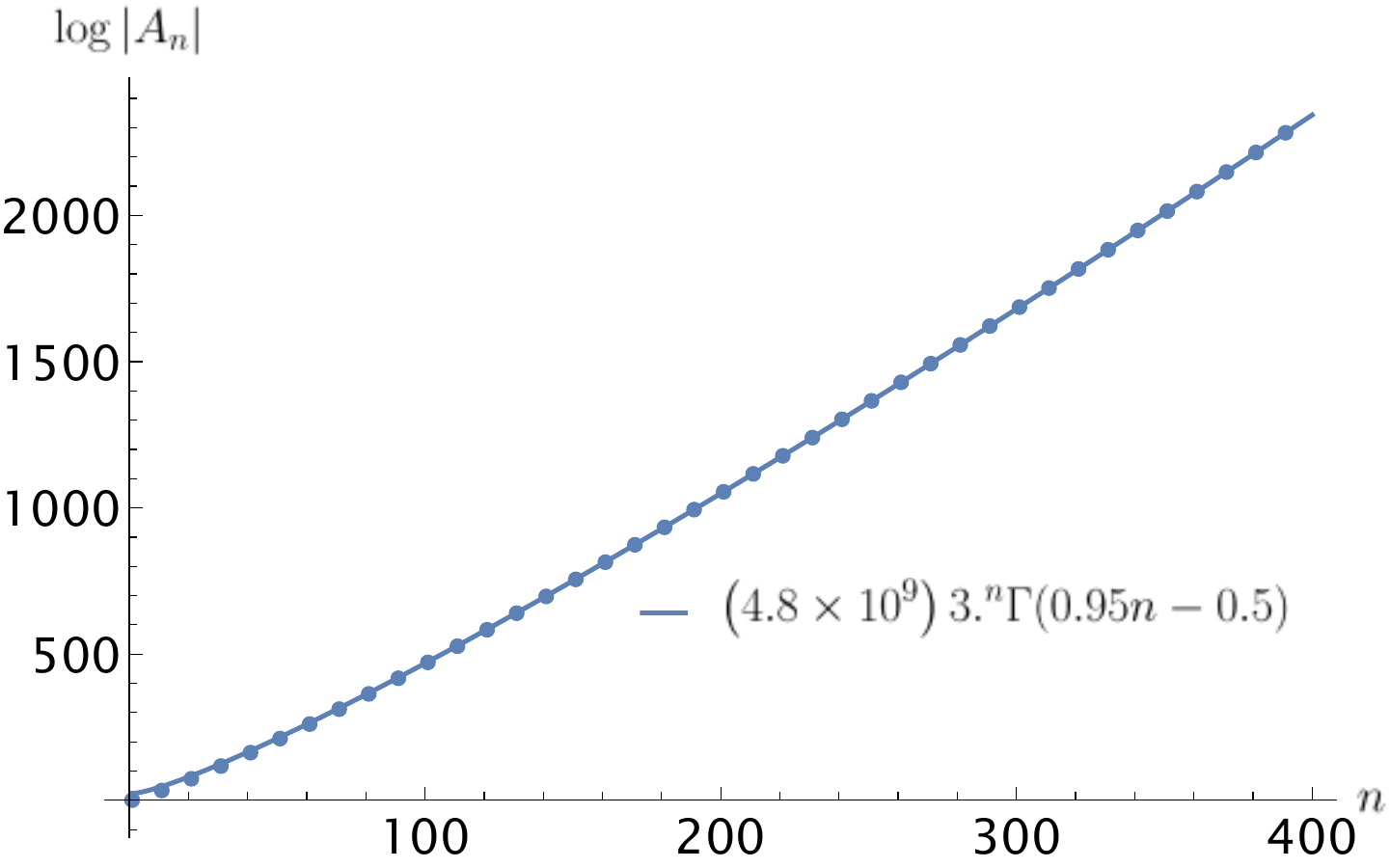}
    \caption{
        The sequence $A_n$ (circles) and a fit using the $\Gamma$ function (solid).
        The sequence $B_n$ has the same behaviour.
    }
    \label{fig:An-fit}
\end{figure}
Calculating the first few coefficients we find
\begin{equation}
    \label{pert-coeffs}
    \begin{split}
        A_k & = \{1, -2, 20, -328, 7024, -179264, \ldots \}, \\
        B_k & = \{1, -6, 80, -1520, 35760, -976704, \ldots \}
        \,
        ,
    \end{split}
\end{equation}
which are seen to alternate in sign and grow quickly.
There are no matching entries in the Online Encyclopedia of Integer Sequences~\cite{oeis}, and we are not aware of a combinatorical interpretation.
Because of the factors $\ell$ appearing inside the sums, the coefficients can be expected to grow factorially; a simple fit confirms graphically that both sequences grow asymptotically as $3^n \Gamma(0.95 n - 0.5)$, see\footnote{While it is known that there is information encoded in such asymptotic growth rates~\cite{Borinsky:2021hnd}, it is out of scope for our present purposes, as we will obtain accurate resummations with only a handful terms.}~\cref{fig:An-fit}.
The series thus has zero radius of convergence.

\begin{figure}[tb]
    \centering
    \subfloat[$\CE = 7.5$]{%
        \includegraphics[width=0.32\linewidth]{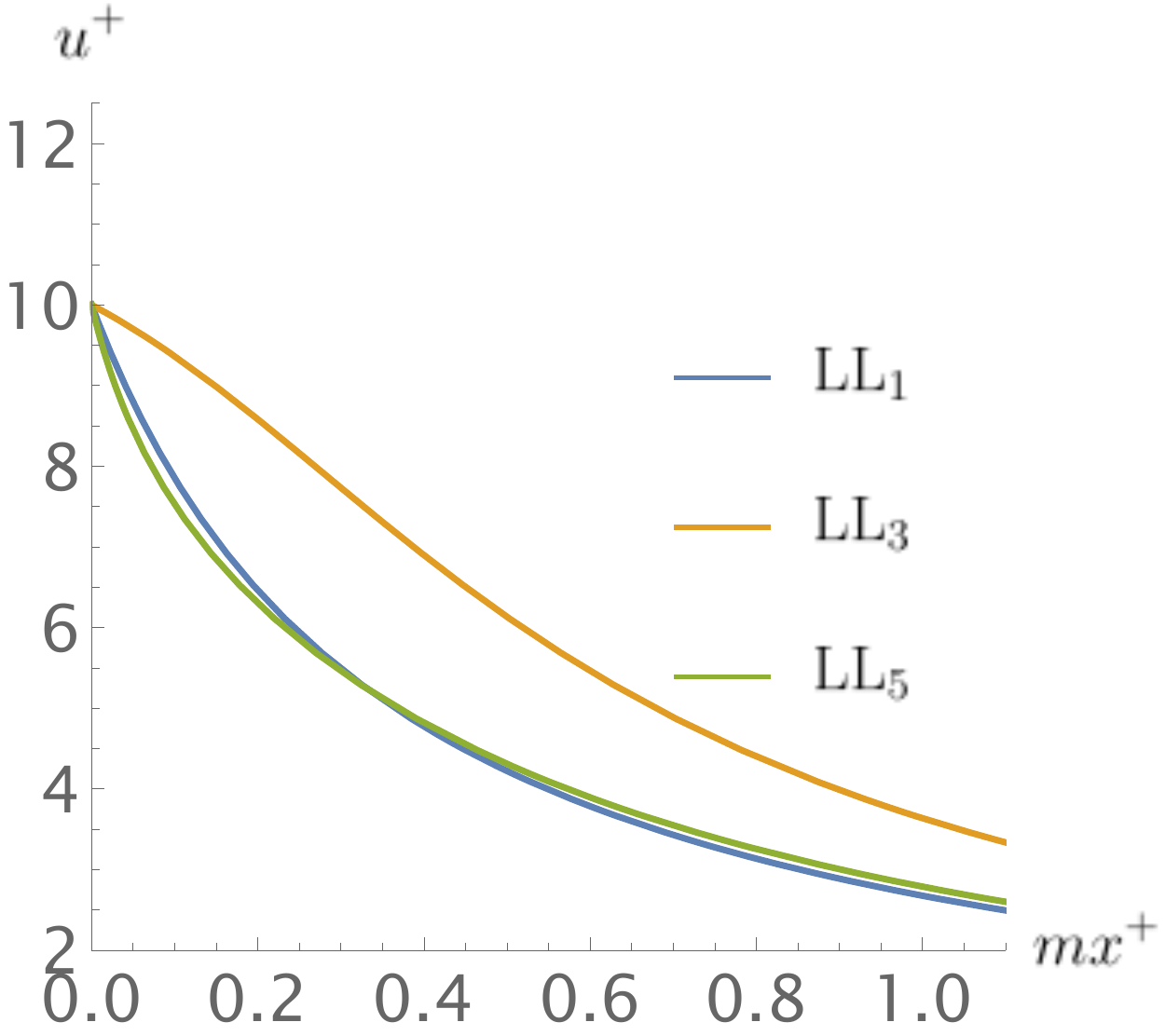}
    }
    \subfloat[$\CE = \frac{1}{\sqrt{6} \tau_0 u_0^\LCp } \approx 8.39$ ]{%
        \includegraphics[width=0.32\linewidth]{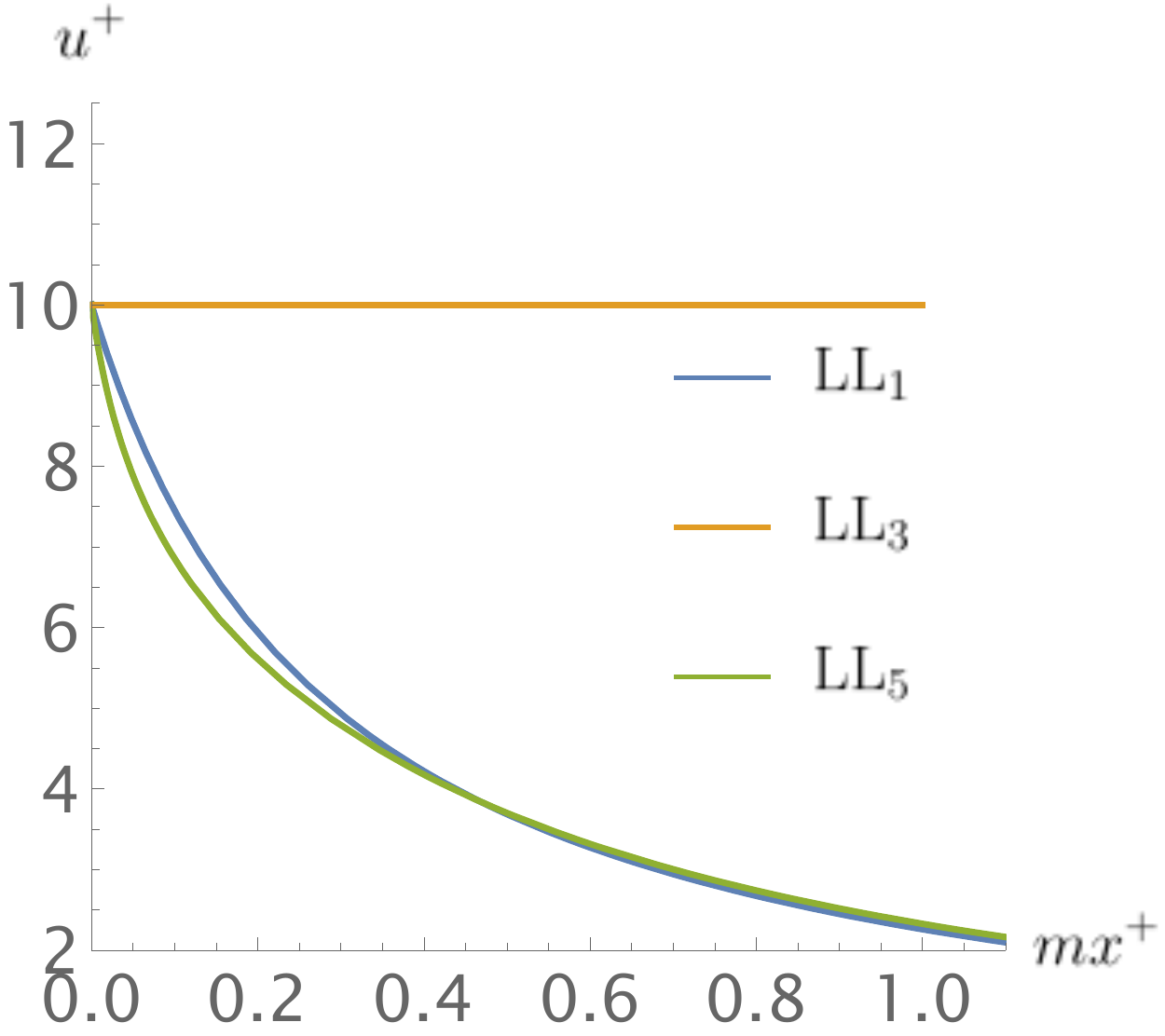}
    }
    \subfloat[$\CE = 10$]{%
        \includegraphics[width=0.32\linewidth]{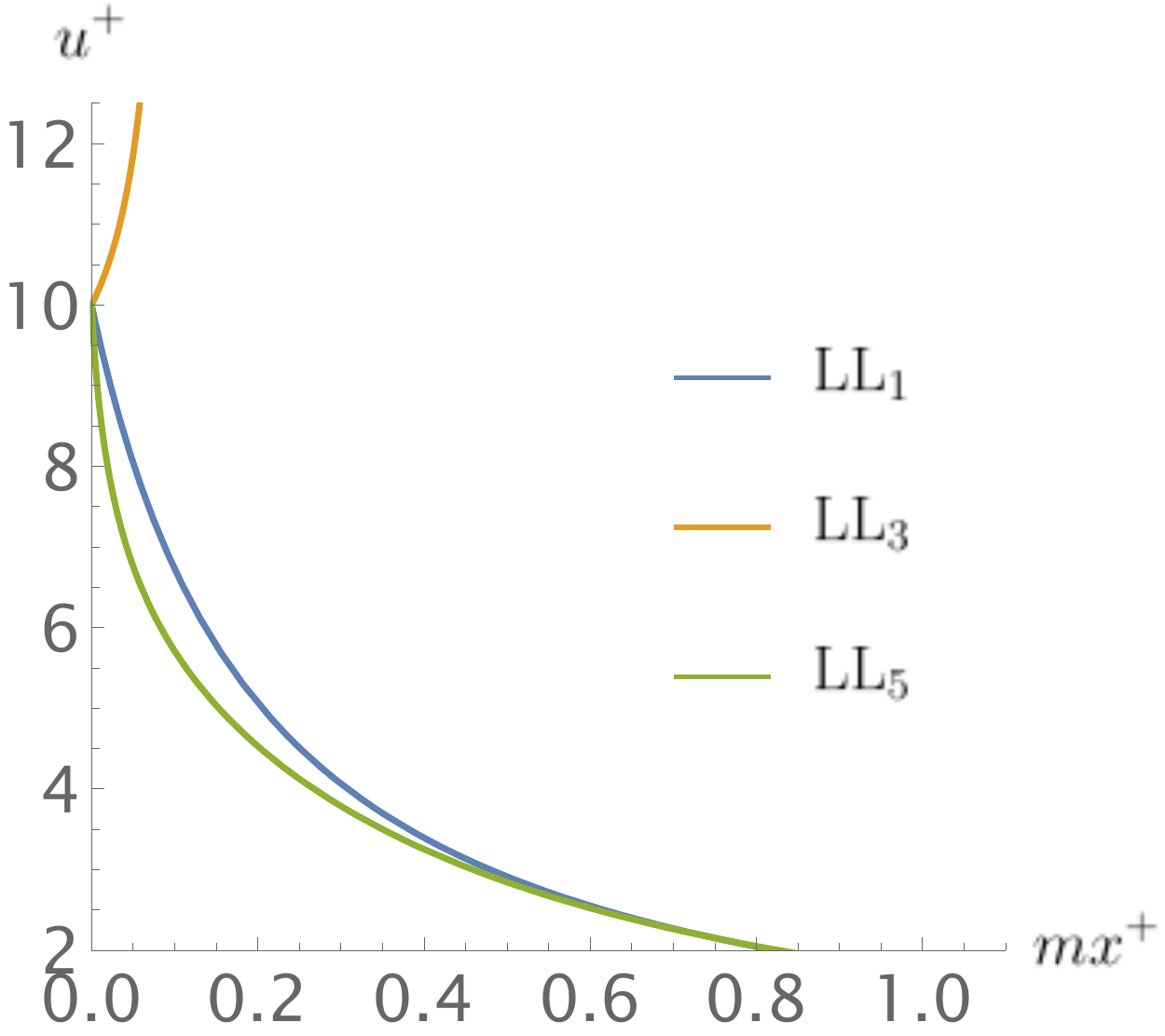}
    }
    \caption{
        Lightfront momentum according to the $\LL1$, $\LL3$, and $\LL5$ equations of motion.
        For a critical value of $\chi$, \LL3 has a stationary solution; for larger $\chi$ this becomes a runaway.
        $\LL5$ is quantitatively wrong in overpredicting the strength of RR at early times, before $\chi$ becomes small.
    }
    \label{fig:LL1-3-5}
\end{figure}

To highlight the physical implications of the divergence of perturbation theory, we present some explicit solutions to the equations of motion $\LL{k}$ for low $k > 1$.
To do so we use lightfront coordinates $x^\LCpm := x^0 \pm x^3, x^\LCperp = \{x^1, x^2\}$, choosing $n\cdot x = x^\LCp$, lightfront time, and $\chi = \CE u^\LCp$.
The $\LL{k}$ equations of motion for $u^\LCp$ and $u^\LCperp$ decouple for all $k$, and become, for any approximation to the functions $\CA$ and $\CB$,
\begin{align}
    \label{eq:LLinf-plus}
    \frac{\ud u^\LCp}{\ud x^\LCp} & = - \tau_0 \chi^2
    {\CB(\chi)} \;, \qquad
    u^\LCp \frac{\ud}{\ud x^\LCp} {\frac{u^\LCperp}{u^\LCp}} =
    {-\CA(\chi) \epsilon^\LCperp}
    \,
    ,
\end{align}
with $u^\LCm$ determined by the mass-shell condition.
The $u^\perp$ components are determined by quadrature once $u^\LCp$ is known, so we will focus on determining $u^\LCp$ in what follows.
The equation in~\cref{eq:LLinf-plus} for $u^\LCp$ is separable, with solution
\begin{equation}
    \label{eq:uplus-sol}
    \tau_0 \int_0^{x^\LCp} \ud y^\LCp
    = \tau_0 x^\LCp
    = -\int_{u^\LCp_0}^{u^\LCp} \frac{\ud v}{\CE^2 v^2 \CB(\CE v)}
    \,
    .
\end{equation}
This \emph{causal} integral can be performed as (for $\CB$ calculated in perturbation theory) the integrand is a rational function.
At first and third order this yields
\begin{align}
    \label{eq:uplus-implicit}
    \LL1: & \quad
    u^\LCp  = \frac{u^\LCp_0}{1 + \tau_0 \CE^2 x^\LCp} \\
    \LL3: & \quad
        \frac{1}{\CE^2 v}
      - \frac{\sqrt{6} \tau_0}{\CE v} \operatorname{arctanh} \sqrt{6} \CE \tau_0 v
      \Big|_{v = u^\LCp_0}^{v = u^\LCp}
    = \tau_0 x^\LCp
    \;
    ,
\end{align}
with \LL5 admitting a similar, but unwieldy and unenlightening expression, which we omit.
Note that, even at only third order, we just obtain an implicit expression for $u^\LCp$.
We therefore proceed graphically, showing solutions to $\LL1$, $\LL3$, and $\LL5$ in \cref{fig:LL1-3-5}.
The features of these solutions can be read off from the respective series expansion of $\CB$.
Recall first that $u^\LCp$ is conserved according to the Lorentz force equation, but not according to $\LL1$.
As $\CB^{(3)}(\chi) = 1 - 6(\tau_0 \chi)^2 < 1$, $\LL3$ predicts less RR than $\LL1$, up until a stationary solution where $\CB^{(3)}$ crosses zero and hence $\chi$, therefore $u^\LCp$, is again conserved.
For $\chi$ above the zero-crossing $\LL3$ gives an RR force in the `wrong' direction, hence predicting a runaway $u^\LCp$ which goes to infinity.
On the other hand $\LL5$ eventually predicts stronger RR than $\LL1$, as $\CB^{(5)} \to +\infty$ for large $\chi$.
This is quantitatively inconsistent with LAD; it has been shown that, writing $u_0^\LCp$ for the initial velocity, $u_0^\LCp \ge u^\LCp_\mathrm{LAD} \ge u^\LCp_\mathrm{\LL{1}}$~\cite{Kazinski:2013vga}, i.e.~\LL1 overestimates RR compared to LAD~\cite{Kazinski:2010ce}.
So, whenever $u^\LCp_\mathrm{\LL{n}} < u^\LCp_\mathrm{\LL{1}}$, the latter is the better approximation.

The conclusions for $\LL3$ and $\LL5$ generalise: as the $\CB^{(n)}$ alternate between diverging to $\pm \infty$, solutions to $\LL{n}$ will be radically different from order to order.
For orders without a runaway, quantitative agreement with LAD becomes worse and worse as $\CB^{(n)}$ grows more rapidly with $\chi$.
We speculate that some of these features may be related to a bifurcation phenomenon associated with LAD in plane wave fields which has recently been discovered in the non-relativistic limit\footnote{{For simple potential steps similar observations go back to~\cite{Bopp:1943} and~\cite{Haag:1955}.}}~\cite{Carati:2021pqd}.
In the absence of any fully relativistic solution to LAD in a plane wave, we try to extract physical results from the diverging series~\cref{eq:LLn} by resumming them.
We will thus obtain an equation of motion, $\LL\infty$, which is free of the problems of $\LL{k}$.

\subsection{Resummation}\label{sec:resumpert}
We note that the series~\cref{eq:LLn} with coefficients~\cref{pert-coeffs} are, although divergent, Borel summable.
This prompts two questions.
First, is there an optimal order of truncation (an optimal $\LL{k}$), as is typical of asymptotic series?
Second, what insights can be gained from resumming the series? We investigate the second question here, returning to the first later on.
We will use the Borel-Pad\'e~\cite{Kleinert:2001ax,Caliceti:2007ra,Costin:2020hwg} method to resum the series $A_k, B_k$.
While there are other resummation methods of potentially higher accuracy, such as Meijer-$G$ resummation~\cite{Mera:2018qte}, Borel-Pad\'e has been successfully applied to several topics in QED~\cite{Florio:2019hzn,Torgrimsson:2020mto,Torgrimsson:2020wlz,Torgrimsson:2021wcj,Dunne:2021acr}, is comparatively simple to implement, and we can easily generate many terms, should they be needed.
(We do not expect any instabilities related to nonperturbative imaginary parts typical for non-alternating coefficients~\cite{Dunne:2002rq,Heinzl:2006pn}.)

\begin{figure*}[tbp!]
    \centering
    \subfloat[\label{fig:borel:a}]{%
        \includegraphics[width=0.48\linewidth]{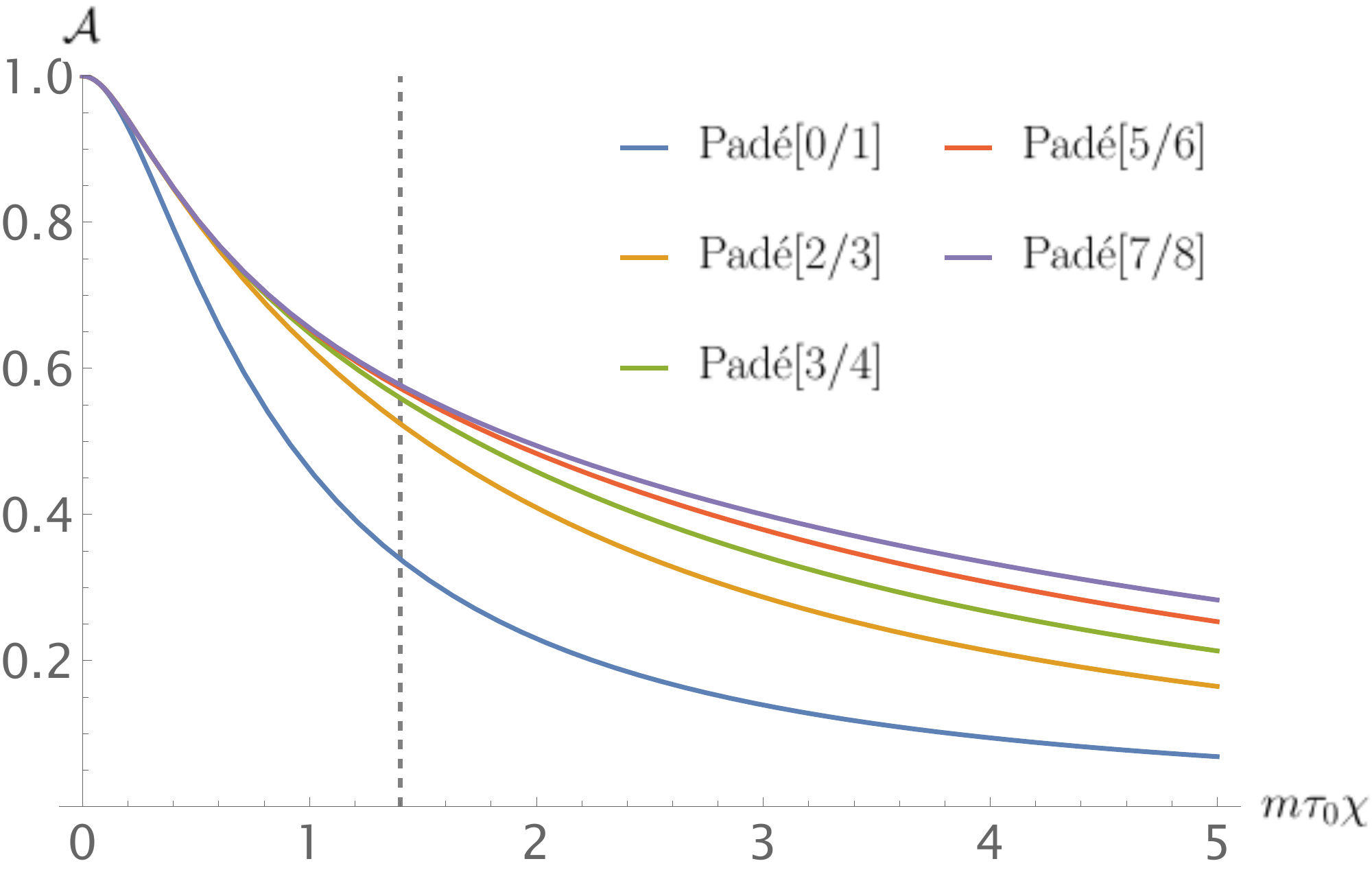}
    }
    \subfloat[\label{fig:borel:a-zoom}]{%
        \includegraphics[width=0.48\linewidth]{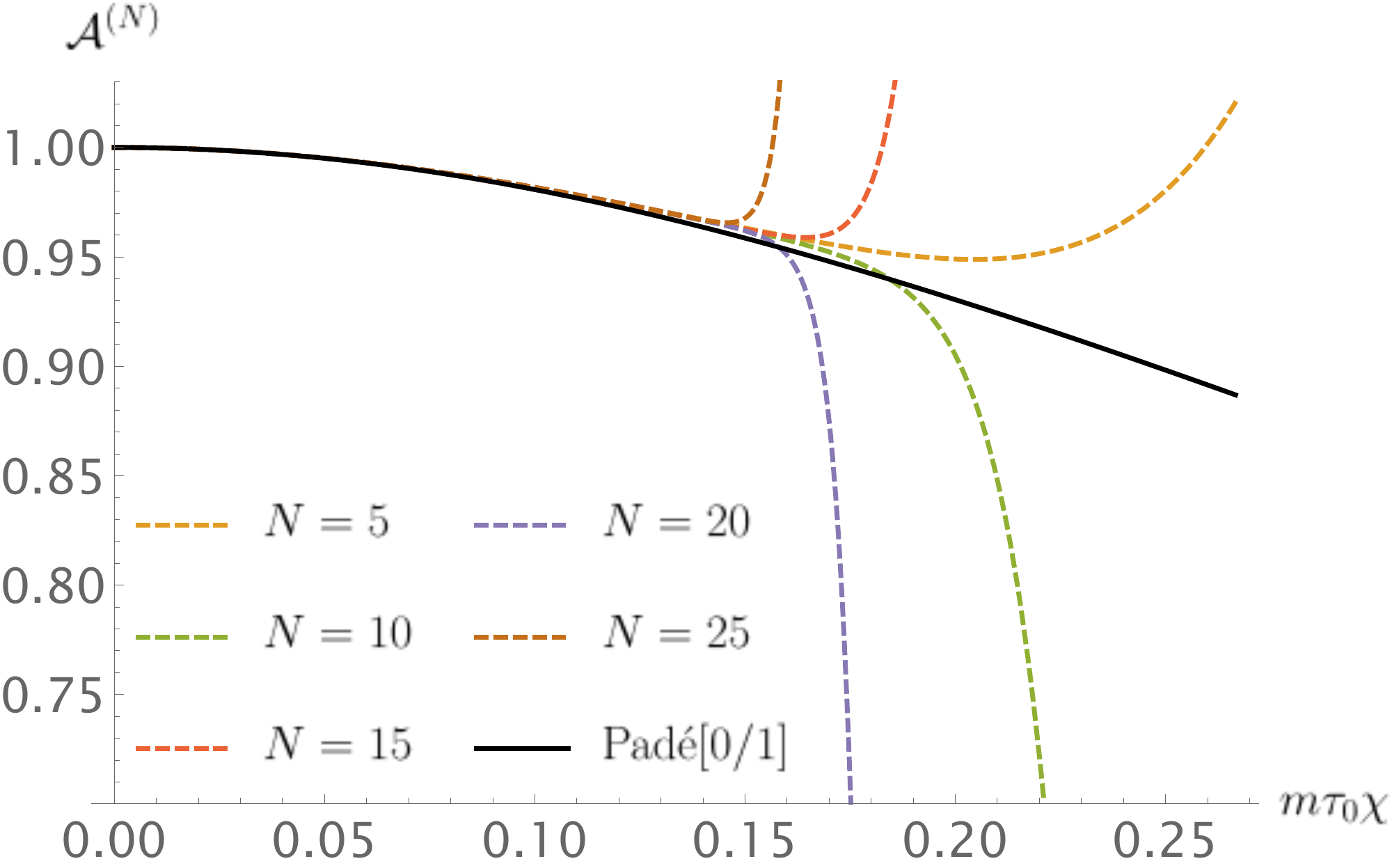}
    }

    \subfloat[\label{fig:borel:b}]{%
        \includegraphics[width=0.48\linewidth]{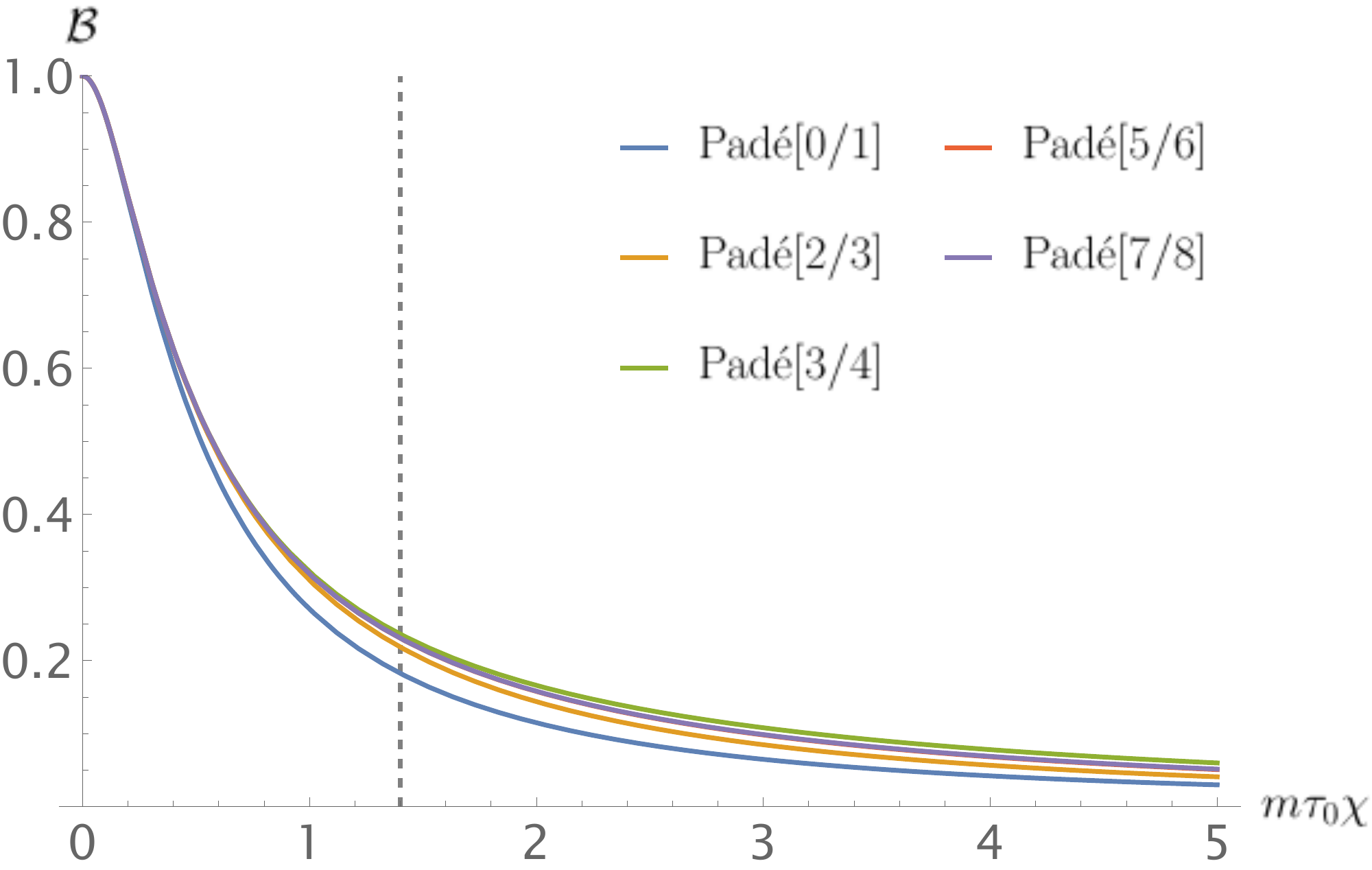}
    }
    \subfloat[\label{fig:borel:b-zoom}]{%
        \includegraphics[width=0.48\linewidth]{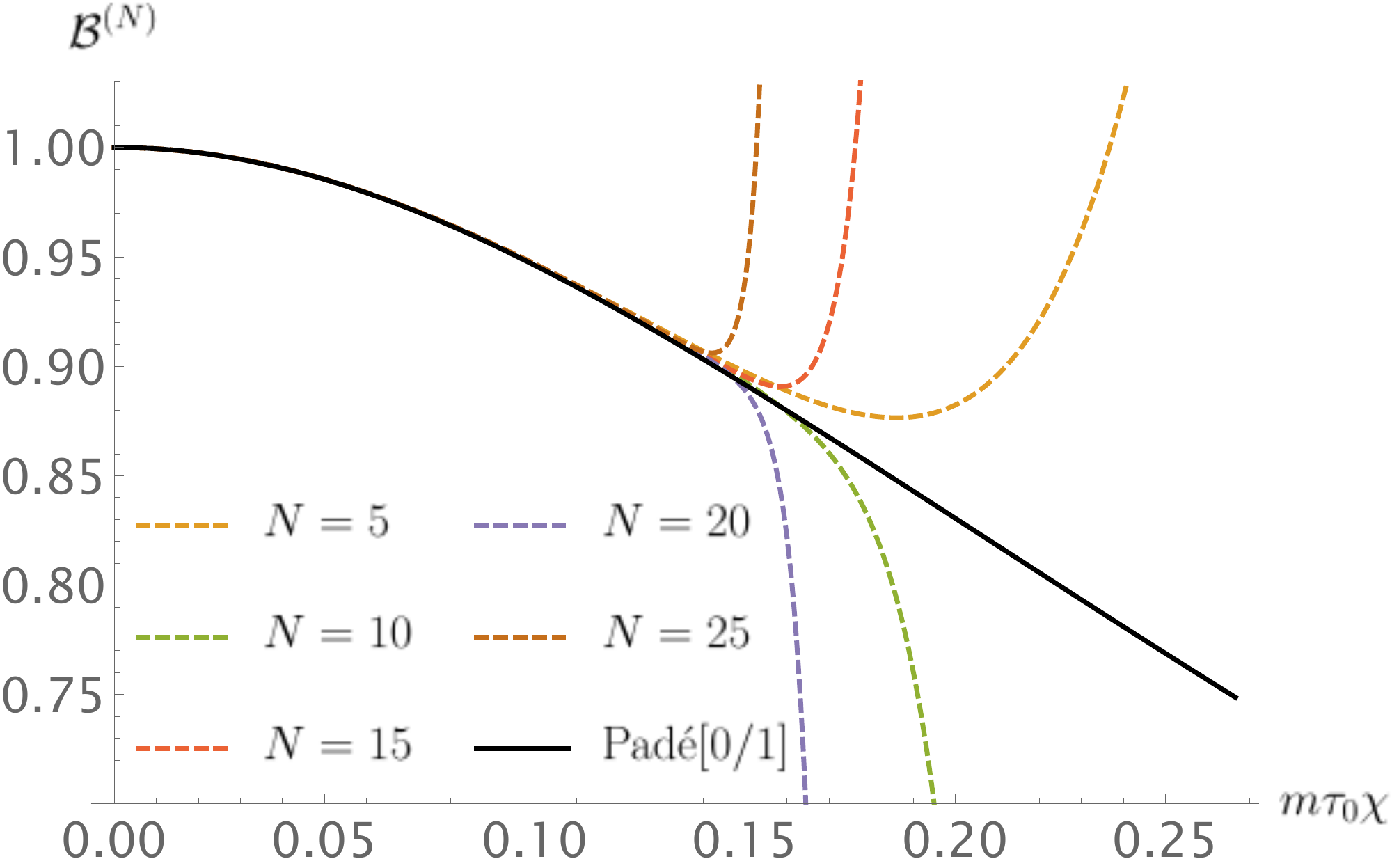}
    }
    \caption{
        Panels \protect\subref{fig:borel:a}, \protect\subref{fig:borel:b}:
        Borel-Pad\'e resummants of the sequences $A_k, B_k$.
        The dashed vertical line marks the radius of convergence of the large-$\chi$ expansion, see \cref{sec:large}.
        Panels \protect\subref{fig:borel:a-zoom}, \protect\subref{fig:borel:b-zoom}:
        Finite-order partial sums (dashed) and the lowest-order resummant (solid black).
        On the interval displayed in these panels one term is sufficient for convergence.
    }
    \label{fig:borels}
\end{figure*}

The method is as follows.
Given $N + M$ perturbative coefficients $A_\ell$ for $\CA$ as a series in $(\tau_0 \chi)^2$, their Borel transform is $\sum A_\ell t^\ell/ \ell!$.
One constructs a Pad\'e approximant of order $N/M$ of the Borel transform,
\begin{equation}
    P_\CA[N/M](t)
    = \frac{\sum_{\ell=0}^N c_\ell t^\ell}{1 + \sum_{j=1}^M d_\ell t^\ell}
    = \sum_{\ell = 0}^{M+N} \frac{A_\ell}{\ell!} t^\ell + \mathcal{O}(t^{M+N+1})
    \,
    ,
\end{equation}
and the resummed series is given by the inverse Borel transform
\begin{equation}
    \mathcal{A}^{[N/M]}(\chi) := \int_0^\infty \ud t \, e^{-t} P_\CA[N/M]\left(t (\tau_0 \chi)^2 \right)
    \,
    ,
\end{equation}
with similar expressions holding for $\CB$.
The rate of convergence can depend on the choice of $N, M$.
We have found $N = M - 1$ to give the fastest convergence (and we will see in the next section why this is).
However, we stress that other choices converge to the same functions, just requiring more terms to do so.
The lowest-order resummants have comparatively simple analytical expressions,
\begin{align}
    \CA^{[0/1]}(\chi) & =
    \frac{e^{\frac{1}{2 \tau_0^2 \chi^2}} \E_1\big( \frac{1}{2 \tau_0^2 \chi^2} \big)}{2 \tau_0^2 \chi^2}
    \quad \xrightarrow{\chi \to \infty} \quad
    \frac{2 \log \tau_0 \chi + \log 2 - \gamma_\text{E} }{2 \tau_0^2 \chi^2}
    \;
    ,
    \label{eq:A01}
    \\
    \CB^{[0/1]}(\chi) & =
    \frac{e^{\frac{1}{6 \tau_0^2 \chi^2}} \E_1\big( \frac{1}{6 \tau_0^2 \chi^2} \big)}{6 \tau_0^2 \chi^2}
    \quad \xrightarrow{\chi \to \infty} \quad
    \frac{2 \log \tau_0 \chi + \log 6 - \gamma_\text{E} }{6 \tau_0^2 \chi^2}
    \;
    ,
    \label{eq:B01}
\end{align}
where $\E_1$ is the exponential integral~\cite[Ch.~5]{AbramowitzandStegun}.
These give, comparing against resummants calculated using more terms, accurate results up $\tau_0 \chi \lesssim 0.3$, as shown in \cref{fig:borels}.
In contrast to the perturbative series the resummants are monotonically decreasing with $\chi$.
The resummed $\LL{n}$ equations of motion therefore predict less RR than $\LL1$, and may have better agreement with LAD.
Notably, since the resummants are always positive, the unphysical runaway solution is eliminated from those $\LL{n}$ that feature it, such as $\LL3$.
Instead, $u^\LCp$ is monotonically decreasing, which means that at large lightfront times $\chi \to 0$; as a result, $\CA, \CB \to 1$, so that the dynamics becomes governed by $\LL1$.

The right-hand panels of \cref{fig:borels} show that for a given $\chi$ the perturbative series agrees with the resummants when no more than $\sim \frac{1}{2} (\tau_0 \chi)^{-2}$ terms are included.
This is in line with the heuristic rule for asymptotic series, that the optimal truncation order is after the smallest term, which occurs at order inversely proportional to the expansion parameter~\cite{Boyd1999}.
As the coefficients defined by~\cref{eq:AkBk} grow factorially from the outset, the optimum order of truncation is one, i.e.~$\LL1$ for $\tau_0 \chi \gtrsim 0.3$.
This suggests that, given their much more complicated form and $\CA, \CB$ not differing greatly from $unity$ for small $\chi$, in practice no finite-order $\LL{k}$, $k \ge 1$, is `better' than $\LL1$.
In the following we will examine the resummed, all-orders equation of motion.
(We will see that the asymptotic logarithms in~\cref{eq:A01} and~\cref{eq:B01} are artefacts of the resummation procedure, but the qualitative features are as described here.)

The resummation above should be contrasted with that in the non-relativistic limit.
In that case the LAD equation is linear in the field strength, resummation only involves derivative terms, and turns out to be straightforward.
It can be used to recover the pre-acceleration solution of LAD from perturbative, reduction of order approximations~\cite{Zhang:2013ria}.

\section{Landau-Lifshitz equation to all orders}
\label{sec:LL-to-all-orders}
\begin{figure}[tb]
    \centering
    \includegraphics[width=0.5\linewidth]{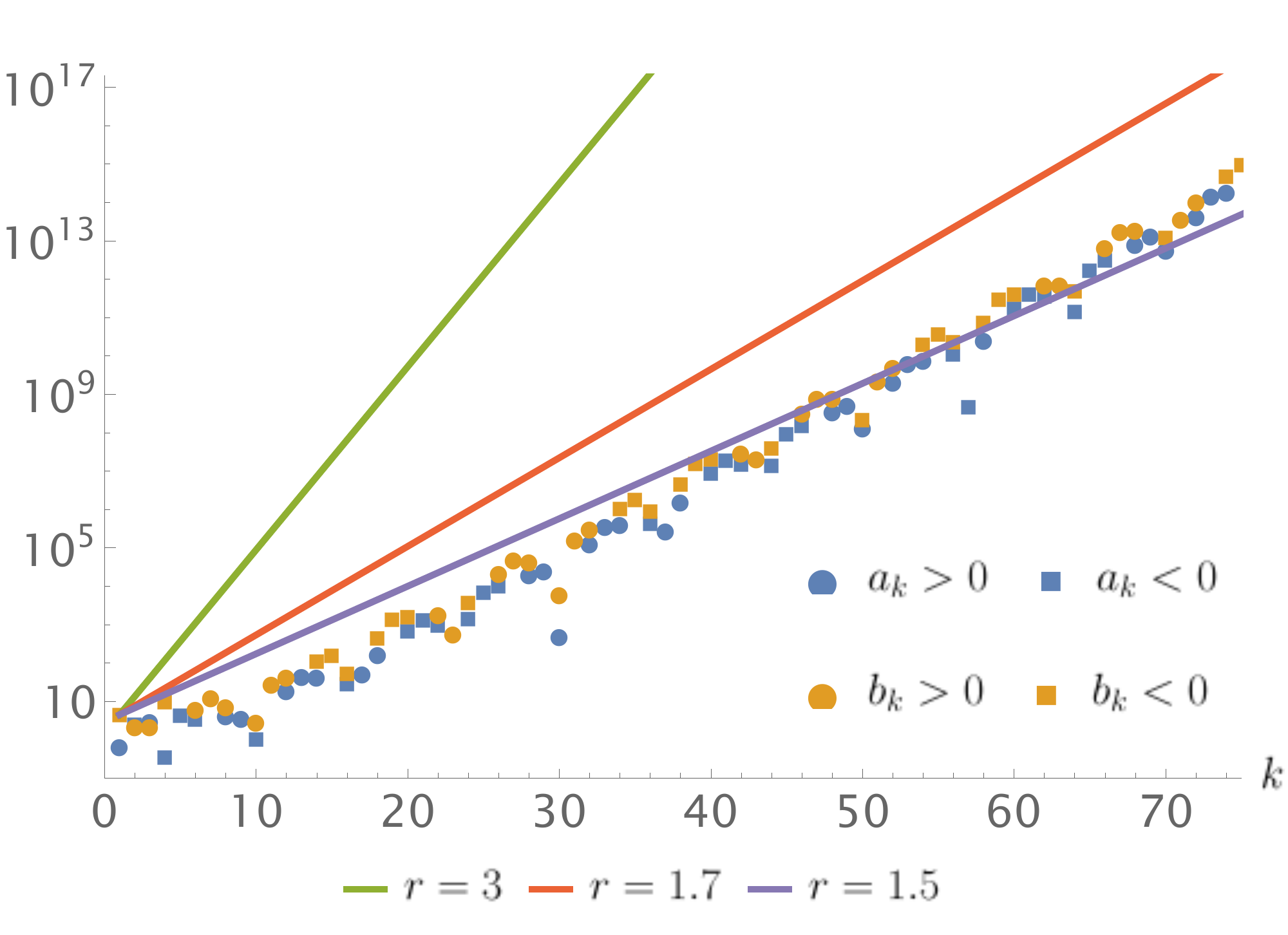}
    \caption{
        The first few terms $a_k$ (blue) and $b_k$ (orange) of the series~\cref{eq:c-series} and trendlines $|b_1| r^{k-1}$ (solid lines).
        Trend lines that lie above all data points give an estimate for the radius of convergence.
        The elements of the sequences have essentially random signs; to resolve these on a log scale, circles (squares) indicate positive (negative) values.
    }
    \label{fig:akbk-bounded}
\end{figure}

Consider now the equation~\cref{GENSOL} obtained by iterating reduction of order infinitely many times, i.e.~including~\cref{eq:LLn} to all orders in $\tau_0$.
We call this equation $\LL\infty$.
By demanding that this be a fixed point of reduction of order, or equivalently by demanding that $u^\mu$ in~\cref{GENSOL} obeys LAD, one finds that $\CA$ and $\CB$ are fully defined by the initial value problem
\begin{equation}
    \label{CAochCB}
    \left\{
    \begin{array}{rcl}
    \tau_0^2 \chi^3 \CB(\chi) \CA'(\chi) &=& 1 - \CA(\chi) - 2 \tau_0^2 \chi^2 \CA(\chi) \CB(\chi) \;, \\ [10pt]
    \tau_0^2 \chi^3 \CB(\chi) \CB'(\chi) &=&   - \CB(\chi) - 2 \tau_0^2 \chi^2 \CB(\chi)^2 + \CA(\chi)^2 \;, \\ [10pt]
    \CA(0) &=& \CB(0) = 1 \; .
    \end{array} \right.
\end{equation}
The solution of these ODEs (which, see below, is easily found numerically) determines $\LL{\infty}$ and its associated dynamics.
There is a subtle sense in which this dynamics must be equivalent to that in LAD, and yet cannot be -- on the one hand,~\cref{CAochCB} is defined by a fixed point condition matching it to LAD,
but on the other hand $\LL\infty$ is second-order in time derivatives and so must be free from pre-acceleration.
While it is clearly not possible to analyse `pre-acceleration before the field turns on’ in our constant field setup, we can easily see that $\LL\infty$ is causal;
the general form¬\cref{eq:uplus-sol} expresses the particle velocity as an integral over past, not future times.
It appears, then, that $\LL\infty$ may match LAD to all-orders in $\tau_0$, but misses non-perturbative effects in $\tau_0$ (runaways and pre-acceleration), the recovery of which would likely require trans-series resummation~\cite{Dunne:2012ae}.
We will compare to LAD below, but first we investigate the properties of $\LL\infty$ in more detail.

The dynamics of $\LL\infty$ has the same \emph{qualitative} features as the dynamics of the resummed $\LL{n}$:
$\CA, \CB$ are monotonically decreasing and as $\CB \le 1$, the RR force is no stronger than that of $\LL1$;
as $\CB$ is always positive, $u^\LCp$ and hence $\chi$ are always driven to zero.
The asymptotics of the solutions to $\LL\infty$ and $\LL1$ are therefore the same.
There are however \emph{quantitative} differences at large $\chi$, as we will now see.

\subsection{\texorpdfstring{Large $\chi$ behaviour and convergence}{Large χ behaviour and convergence}}
\label{sec:large}
We cannot solve the fixed-point ODEs analytically, but a series ansatz around $\chi = 0$ reproduces exactly the recursion relations~\cref{eq:AkBk}.
This prompts us to try a (Frobenius) series expansion for large~$\chi$; we find
\begin{align}
        \CA(\chi) = \sum_{k=0}^\infty a_k \Big(\frac{3 \tau_0}{\sqrt{2}} \chi\Big)^{-(k+1)/2}
        \;,
        \qquad
        \CB(\chi) = \sum_{k=0}^\infty b_k \big(\frac{3 \tau_0}{\sqrt{2}} \chi\Big)^{-(k+3)/2}
        \;
        ,
        \label{eq:c-series}
\end{align}
where the factor $3/\sqrt{2}$ has been introduced for convenience.
The leading coefficients are $a_0 = 1$, $b_0 = 3$, with the higher orders determined by the recursion relation
\begin{widetext}
    \begin{equation}
        \label{eq:akbk-recur}
        \frac{1}{3} \begin{pmatrix}
            3 - k & 1 \\
            -6    & 2-k
        \end{pmatrix}
        \begin{pmatrix}
            a_k \\ b_k
        \end{pmatrix} =
        \begin{pmatrix}
            -a_{k-1} + \frac{1}{9} \sum (\ell - 3) a_\ell b_{k-\ell} \\
            -b_{k-1} + \sum a_\ell a_{k-\ell} + \frac{1}{9} \sum (\ell - 1) b_\ell b_{k-\ell}
        \end{pmatrix} \; ,
    \end{equation}
\end{widetext}
the sums running over $1 \le \ell \le k-1$.
Unlike the factorially growing $A_k, B_k$ in~\cref{pert-coeffs}, the coefficients $a_k$, $b_k$ grow no faster than exponentially with $k$, as we illustrate for the first hundred in~\cref{fig:akbk-bounded} and prove in \cref{app:proof}.
The series~\cref{eq:c-series} therefore have finite radii of convergence in $\chi^{-1/2}$; our proof provides a way to estimate the criterion for convergence as
\begin{equation}
    \label{eq:radius-of-convergence}
    \tau_0 \chi \gtrsim 1.37
    \;,
\end{equation}
using the data in \cref{fig:akbk-bounded}.
\begin{figure}[tbp]
    \centering
    \subfloat[]{%
        \includegraphics[width=.49\linewidth]{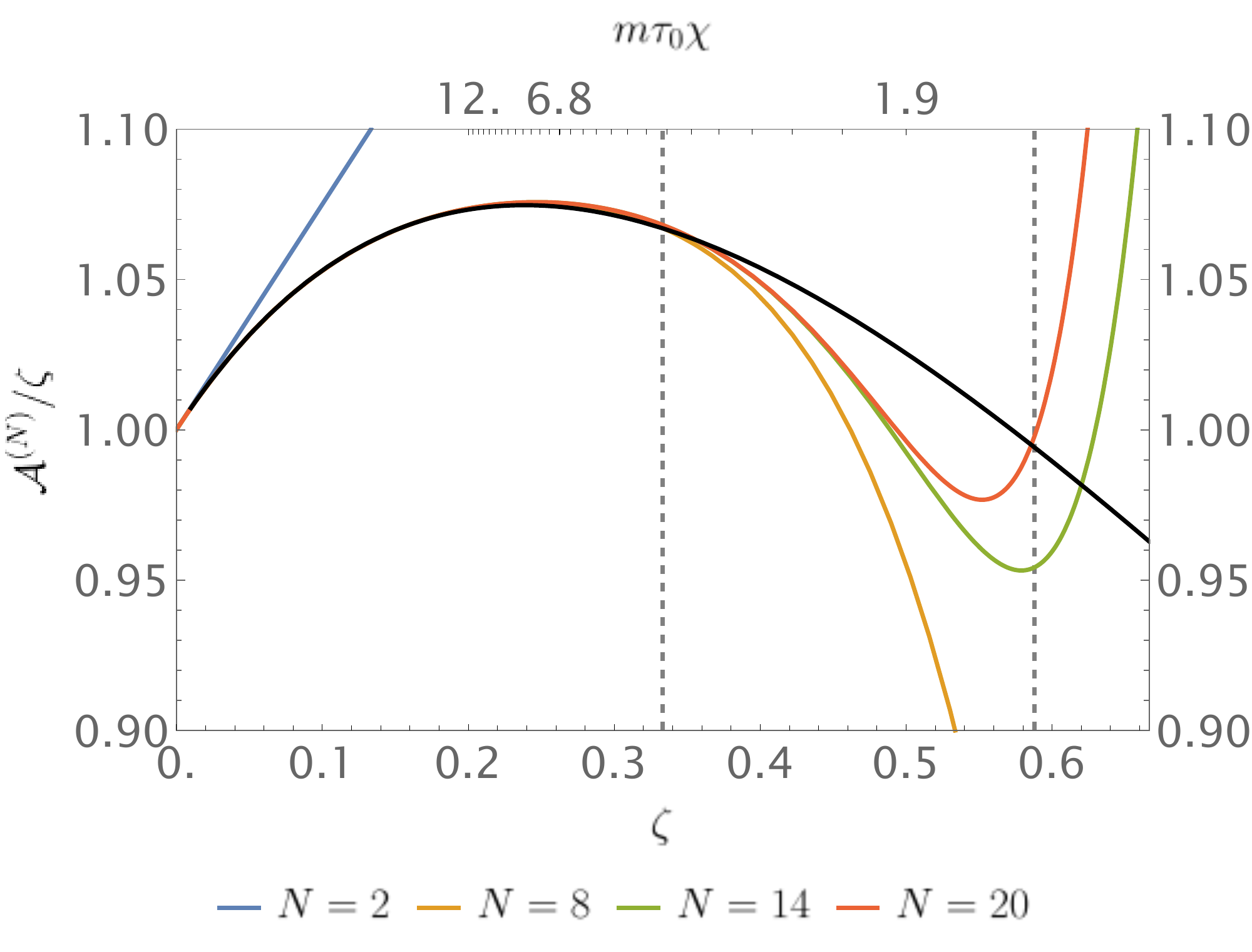}
    }
    \subfloat[]{%
        \includegraphics[width=.49\linewidth]{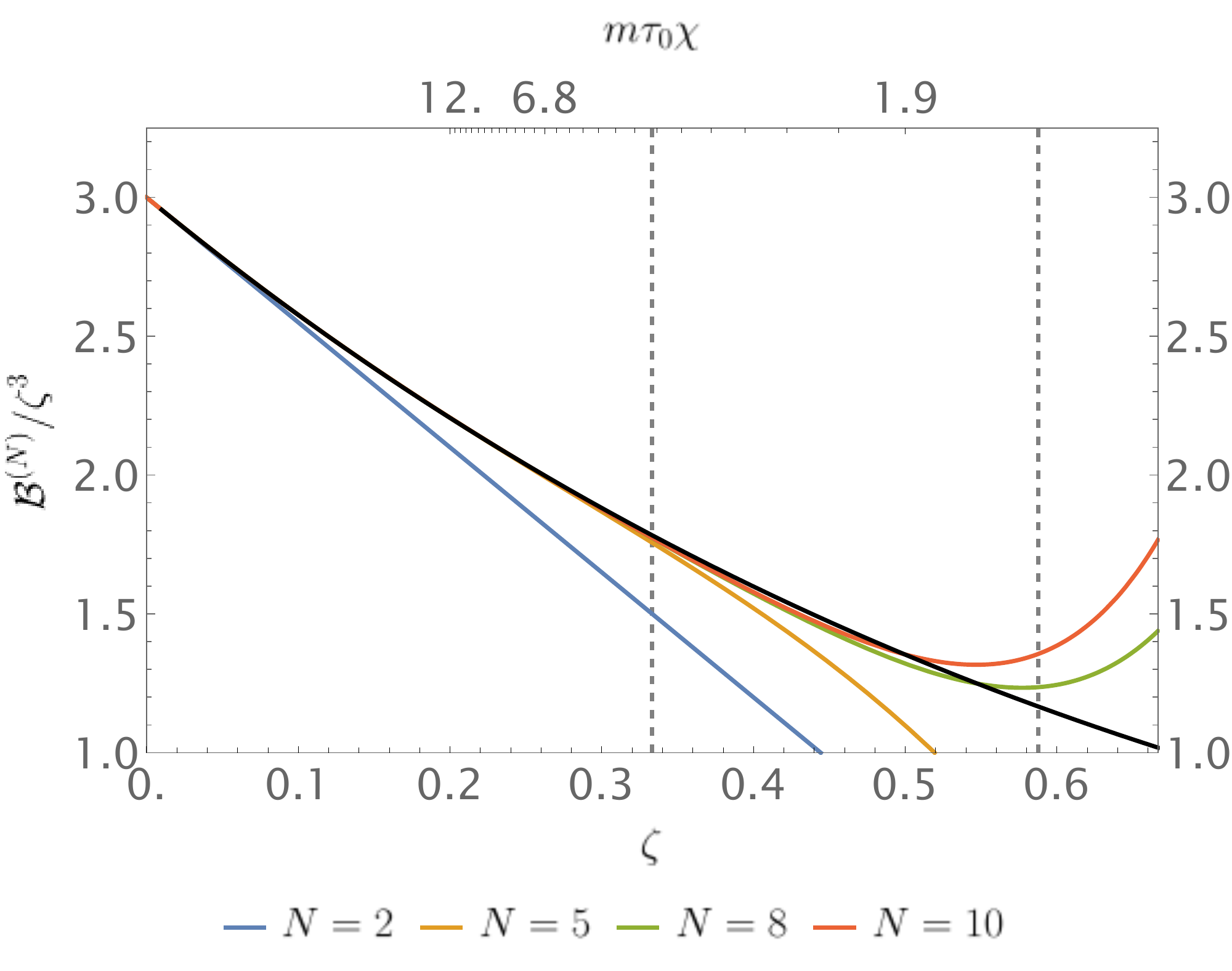}
    }
    \caption{
        Partial sums with $N$ terms for the functions $\CA, \CB$ (colour) and a numerical solution of the fixed-point ODE:s (black).
        The dashed lines indicate unconditional and estimated radii of convergence; see \cref{app:proof}.
        Here $\zeta = (3 \tau_0 \chi/\sqrt{2})^{-1/2}$.
    }
    \label{fig:partial-sums}
\end{figure}
We investigate the convergence of the series~\cref{eq:c-series} in more detail in~\cref{fig:partial-sums}, by comparing against a numerical solution of the equations~\cref{CAochCB} for $\chi$ large; the latter requires setting conditions at $\chi=\infty$, where the fixed-point ODEs are singular, but since the series solution is convergent, we can set an initial condition $\CA\big( ( \frac{3}{\sqrt{2}} \tau_0 \chi)^{-1/2} = \varepsilon\big) = \varepsilon + a_1 \varepsilon^2 + \ldots$, thereby avoiding the singularity.
\cref{fig:partial-sums} shows the partial sums indeed converging to the numerical solution when condition~\cref{eq:radius-of-convergence} holds.
The numerical solution can be extended down to small values of $\chi$, which can also be used to check on the resummation of the perturbative series, see below.

As the series~\cref{eq:c-series} are convergent for large $\chi$, we are encouraged to investigate the leading-order high-$\chi$ dynamics implied by $\LL{\infty}$~\cref{CAochCB}, which is encoded in the $k = 0$ terms of~\cref{eq:c-series}.
These yield the large-$\chi$ asymptotics
\begin{equation} \label{eq:large-chi}
    \CA \; \sim \;  \frac{2^{1/4}}{3^{1/2}} \chi^{-1/2} \; ,
    \quad
    \CB \; \sim \; \frac{2^{3/4}}{3^{1/2}} \chi^{-3/2} \; ,
\end{equation}
in which we note that the powers appearing are the inverse of those typically associated with the Ritus-Narozhny conjecture~\cite{Mironov:2020gbi}.
As before we focus on the $u^\LCp$ component, the equation for which, recall~\cref{eq:LLinf-plus}, here becomes
\begin{equation}
    \label{eq:large-chi-u}
    \int_{u_0^\LCp}^{u^\LCp} \frac{\ud v^\LCp}{2\sqrt{v^\LCp}}
    =
    - \int_0^{x^\LCp}
    \bigg(\frac{m \CE}{3\sqrt{2}\tau_0}\bigg)^{1/2}   \, \ud y^\LCp
    \quad
    \implies
    \quad
    u^\LCp(x^\LCp)
    =
    u_0^\LCp\Big( 1 - \sqrt{ \frac{m \CE}{3u_0^\LCp\sqrt{2} \tau_0} } x^\LCp\Big)^2
    \,
    .
\end{equation}
Note, again, that the solution is explicitly causal: the solution $u(x^\LCp)$ does not sample the field at future times.
The solution shows that from an initial value of $u_0^\LCp$ at $x^\LCp = 0$, the momentum $u^\LCp$ drops to zero after a finite lightfront time, corresponding to the particle being accelerated to the speed of light~\cite{Woodard:2001hi}.
This is qualitatively consistent with the general discussion of $\LL{\infty}$ above, but the result~\cref{eq:large-chi-u} cannot be quantitatively valid for all times because as $u^\LCp$ drops so does $\chi$ and we leave the high-$\chi$ region, hence the regime of validity of our approximation.
We can estimate, directly from~\cref{eq:large-chi-u}, that the particle remains at $\chi \gg 1$ only for
\begin{equation}
    x^\LCp \lesssim \bigg( \frac{u^\LCp_0 \tau_0}{m \CE} \bigg)^{1/2} \;,
\end{equation}
which can be understood as a measure of the time it takes for the initial lightfront energy $mu^\LCp$ to be radiated away in a field of strength $\CE$.
Consequently the differences between $\LL1$ and $\LL\infty$ will be most apparent if high $\chi$ is reached through high \emph{energy}.
Note that observables in QED have very different behaviours depending on whether $\chi$ is made large through high energy or high intensity, which has implications for the Ritus-Narozhny conjecture~\cite{Podszus:2018hnz,Ilderton:2019kqp}.

\section{Improved resummation}
\label{sec:stitched-stuff}

The preceding results demonstrate that we need the full $\chi$-dependence of the functions $\CA$ and $\CB$, even at high intensity where their series expansions are convergent, because the solution of $\LL\infty$ is driven with lightfront time to the low-$\chi$ regime.
In this section we use resummation to obtain an analytic approximation of the functions $\CA$ and $\CB$ which holds for $0\leq \chi < \infty$.

\begin{figure}[tbp]
    \centering
    \includegraphics[width=0.5\linewidth]{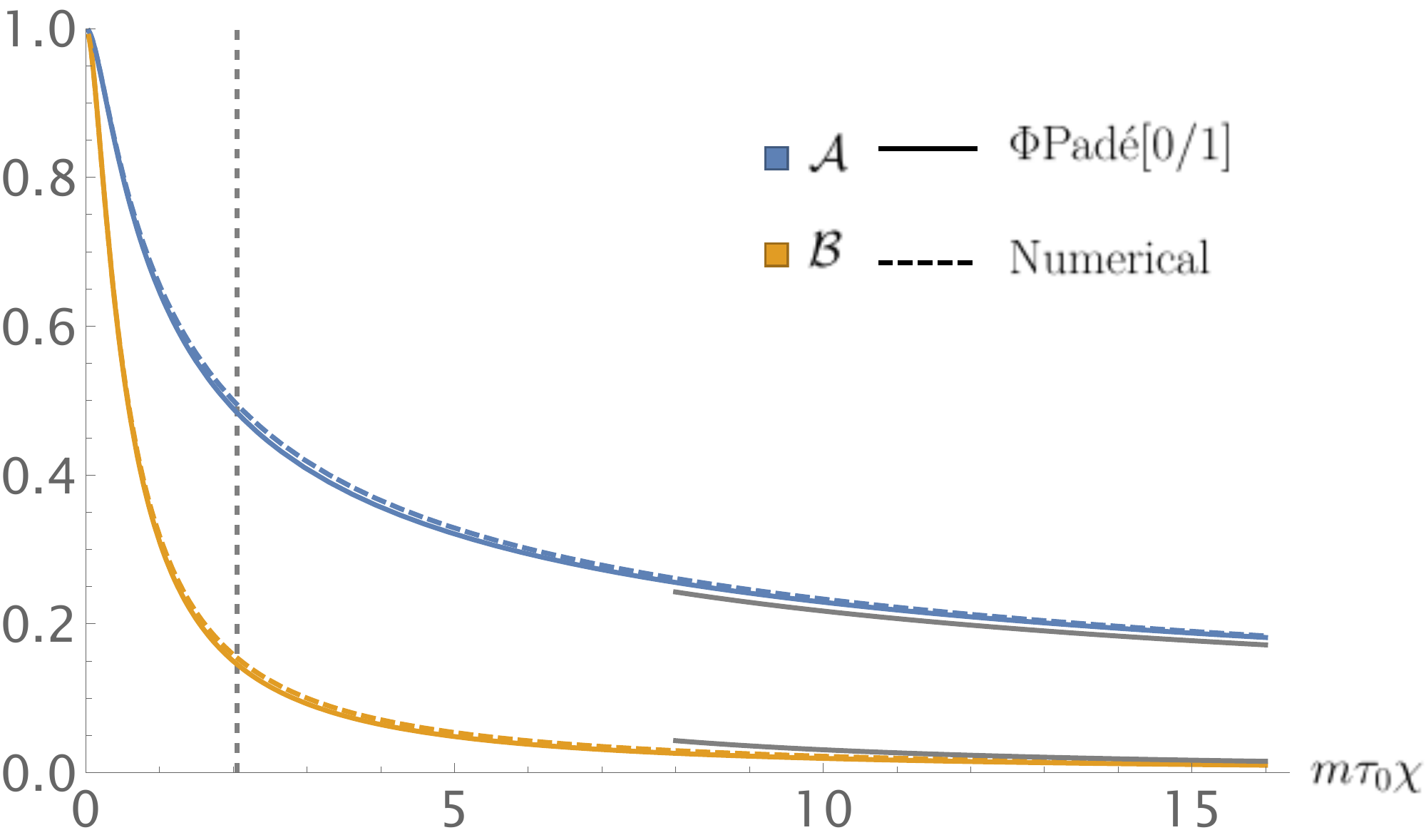}
    \caption{
        Numerical solutions to the fixed-point ODEs (dashed) and the lowest-order $\Phi$-Pad\'e resummant (solid).
        Remarkable agreement is obtained using only a single perturbative coefficient, and by construction the correct asymptotic behaviour (grey curves) is reproduced.
        Vertical dashed grey lines indicate the radius of convergence of the large-$\chi$ series.
    }
    \label{fig:stitching}
\end{figure}

We can see directly from~\cref{eq:large-chi} that the Borel-Pad\'e resummed perturbative series considered in \cref{sec:resumpert}, e.g.~\cref{eq:A01} and~\cref{eq:B01}, do not give the correct high-$\chi$ scaling.
This is due to Pad\'e approximants always having integer power scaling, $N - M$,  for large arguments.
With integer powers, taking $N - M = 1$ is the closest we can get to the correct falloffs~\cref{eq:large-chi}, which is why this choice leads to the most rapid convergence.
We can improve upon this by instead using a resummation method that incorporates asymptotic data;
one such method is ``educated match'' or $\Phi$-Pad\'e resummation~\cite{Alvarez:2017sza}.
In one variant of this method one constructs a Pad\'e approximant
\begin{equation}
    \operatorname{\Phi P}_\CA[N,M](t)
    = \frac{\sum_{\ell=0}^N c_\ell t^\ell}{1 + \sum_{j=1}^M d_\ell t^\ell}
    = \sum \frac{A_\ell t^\ell}{\ell!} \frac{\Gamma(\kappa)}{\Gamma(\kappa + \ell)} \frac{\Gamma(\lambda)}{\Gamma(\lambda + \ell)}
    +
    \mathcal{O}(\chi^{N+M+1})
    \;
    ,
\end{equation}
where the Borel transform has been replaced by a ``hypergeometric transform'' depending on two parameters $\kappa$ and $\lambda$.
The resummant is obtained from the integral
\begin{equation}
    \CA(\chi) = \frac{1}{\Gamma(\kappa) \Gamma(\lambda)} \int_0^\infty \ud t \,
    e^{-t} t^{\kappa - 1} \operatorname{\Phi P}\left(t (\tau_0 \chi)^2\right) U(1 - \lambda, \kappa + \lambda + 1, t)
    \;
    ,
\end{equation}
where $U$ is a confluent hypergeometric function~\cite[Ch.~13]{AbramowitzandStegun}.
The asymptotic behaviour of the resummant is
\begin{equation}
    \CA \xrightarrow{\chi \to \infty}
        (\tau_0 \chi)^{-2 \lambda} \frac{\Gamma(\kappa - \lambda)}{\Gamma(\kappa)}
        + (\tau_0 \chi)^{-2 \kappa} \frac{\Gamma(\lambda - \kappa)}{\Gamma(\lambda)}
        \;
        ,
\end{equation}
from which $\kappa, \lambda$ can be chosen to match a known asymptote.
As with the order of the Pad\'e approximant, different choices affect the rate of convergence.
A convenient choice is $\lambda = 1$, as then one can make the replacement $U \mapsto 1$.
The integral can then be performed analytically for the $[0, 1]$ approximants, superseding~\cref{eq:A01} and~\cref{eq:B01},
\begin{align}
    \label{eq:A01.Phi}
    \CA^{[0/1]}(\chi)
    & =
    \frac{e^{\frac{1}{8 \tau_0^2 \chi^2}}}{8 \tau_0^2 \chi^2}
    \E_{\frac{1}{4}}\left(\frac{1}{8\tau_0^2 \chi^2}\right)
    \quad \xrightarrow{\chi \to \infty} \quad
    0.73 (\tau_0 \chi)^{-1/2}
    \\
    \label{eq:B01.Phi}
    \CB^{[0/1]}(\chi)
    & =
    \frac{e^{\frac{1}{8 \tau_0^2 \chi^2}}}{8 \tau_0^2 \chi^2}
    \E_{\frac{3}{4}}\left(\frac{1}{8\tau_0^2 \chi^2}\right)
    \quad \xrightarrow{\chi \to \infty} \quad
    0.76 (\tau_0 \chi)^{-3/2}
    \,
    ,
\end{align}
where again $\operatorname{E}_n$ is an exponential integral~\cite[Ch.~5]{AbramowitzandStegun}.
These have the correct asymptotic powers of $\chi$ by construction, with coefficients that are notably close to the correct values $2^{1/4}/\sqrt{3} \simeq 0.69$ and $2^{3/4}/\sqrt{3} \simeq 0.97$, despite us having used only a \emph{single} perturbative coefficient to construct this lowest-order resummant.
(Matching the coefficients exactly would require solving a transcendental equation  for $\lambda$, leading to an integral we cannot perform analytically.)
The analytic results~\cref{eq:A01.Phi} and~\cref{eq:B01.Phi} represent an excellent approximation to the numerically exact solution of~\cref{CAochCB} across the whole range of $\chi$, as we demonstrate in~\cref{fig:stitching}.
With this, we turn to a direct comparison of the dynamics implied by $\LL\infty$ with those implied by LAD.
(For comparisons with $\LL1$, see Refs.~\cite{Hadad:2010mt,bulanov2011lorentz,bulanov2017charged,Seipt:2019dnn}.)

\section{Beyond constant fields: comparison with LAD}

Consider now $\LL\infty$ for fields which are not constant, but which vary on time scales much longer than $\tau_0$.
For such fields the derivative terms neglected in the derivation of $\LL\infty$ are small.
We show in \cref{app:derivatives} that our methods can be applied, and our conclusions remain robust, at leading order in derivatives.
We therefore proceed to analyse motion in non-constant fields using a `locally constant field approximation'~\cite{Ritus1985a}, in which we simply promote $f$ and $\chi$ appearing in $\LL\infty$ to lightfront-time dependent variables.

Our test case is a particle created by some mechanism (e.g.~the non-linear Breit-Wheeler process), in the peak of a circularly polarised pulse with a $\sin^2$ envelope and finite duration; the fieldstrength is $f_{\mu\nu} = n_\mu a'_\nu - a'_\mu n_\nu$, where
\begin{equation}
    \label{eq:pulse}
    a'^\LCperp(\phi) =
    \CE \sin^2 \frac{\phi}{4} \{ \cos \phi, \sin \phi \} \; ,
    \quad
    0 \le \phi \le 4 \pi
    \,
    ,
    \qquad \text{ and zero otherwise.}
\end{equation}
We consider this situation, rather than a particle already present before the pulse arrives, for the following reason.
In a strong \emph{pulsed} field, the large-$\chi$ regime where we expect differences between the equations of motion to become sizeable is never reached; even if the \emph{peak} field and \emph{initial} $u^\LCp$ are taken to be large, almost all of the particle's lightfront energy will be radiated away long before it reaches the strong-field region~\cite{Blackburn:2019rfv,Ekman:2021vwg}.
Our scenario thus mimics the situation where a charged particle encounters a `step' at a given phase $\phi_i$ where the field is suddenly switched on.

In the differential formulation~\cref{eq:LAD} of LAD, an additional boundary condition for $\dot{u}^\mu$ is needed, which can
in principle be the physically motivated \emph{final} condition $\dot{u}^\LCp(\phi_f) = 0$, $\phi_f \ge 4\pi$, in our case, i.e.~that the acceleration vanishes after the field has been turned off.
`Numerically exact' solutions of LAD must resolve the particle's proper time on the scale of $\tau_0$ or less, as this multiplies the highest derivative in LAD.
The analytical solution to $\LL1$ implies, though, that the proper time spent in the pulse is of order $3\pi^2 \CE^2 \tau_0$, implying an unreasonable number of time steps for the case of interest, namely high intensities, $\CE^2 \gg 1$.
We will thus use the integro-differential~\cite{Haag:1955,rohrlich1961,Plass:1961zz,PhysRevE.88.033203} formulation of LAD.
It substitutes for the additional boundary condition a pre-accelerating integral over future times of the form
\begin{equation}
    \iota \sim \int_\tau^\infty \ud s \, e^{-(s - \tau)/\tau_0} \mathcal{F}(s)
    \,
    ,
\end{equation}
where $\mathcal{F}$ is a certain function of $u^\mu$ and $f^{\mu\nu}$, see \cref{eq:alcaine} for details;
$\iota$ must be known at $\phi_f = \omega x^\LCp_f$ to solve backwards in time.
The integral vanishes if the field vanishes for all future times, but again we need to resolve proper time on the scale $\tau_0$ and so integrating backwards from $\phi = 4 \pi$ is again unfeasible.
However, after a very short time $\varepsilon$, almost all of the initial lightfront energy present at `creation', $\phi_i = 2\pi$, will have been radiated away, and we can use the solution to $\LL1$ or $\LL\infty$ to compute $\iota(\phi = 2 \pi + \varepsilon)$,
since by this point the particle has reached the regime where all the equations of motion agree.
In any case, the solution is only weakly sensitive to the starting value of $\iota$, as it is exponentially damped on the scale $\tau_0$.
The details of our numerical discretisation scheme are further described in \cref{app:numerics}.

\begin{figure}[tbp]
    \centering
    \subfloat[
        Lightfront component $u^\LCp$.
        \label{fig:LLinf-LAD:plus}
    ]{
        \includegraphics[width=0.48\linewidth]{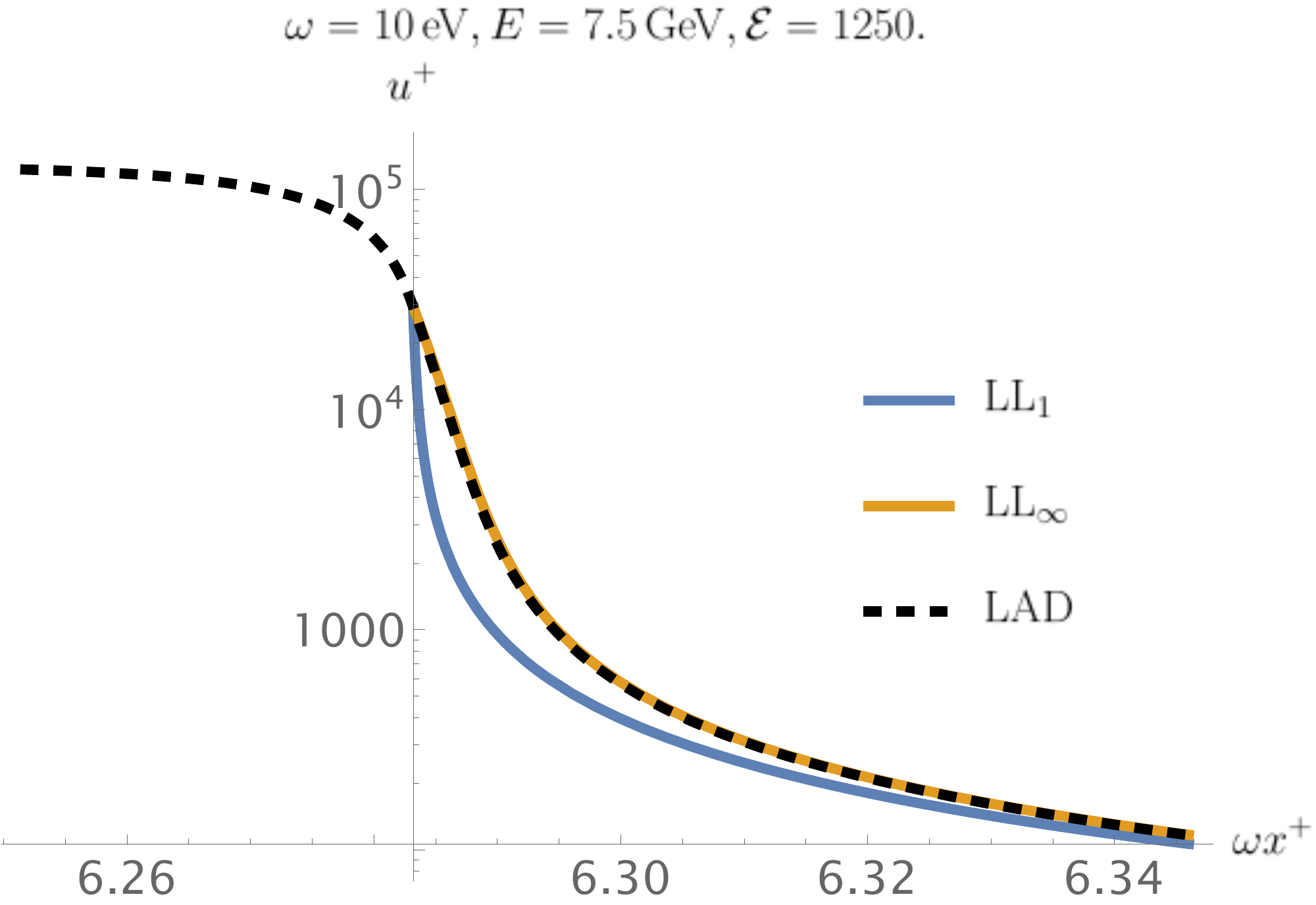}
    }%
    \subfloat[One of the perpendicular components, $u^\mathfrak{2}$.]{
        \includegraphics[width=0.48\linewidth]{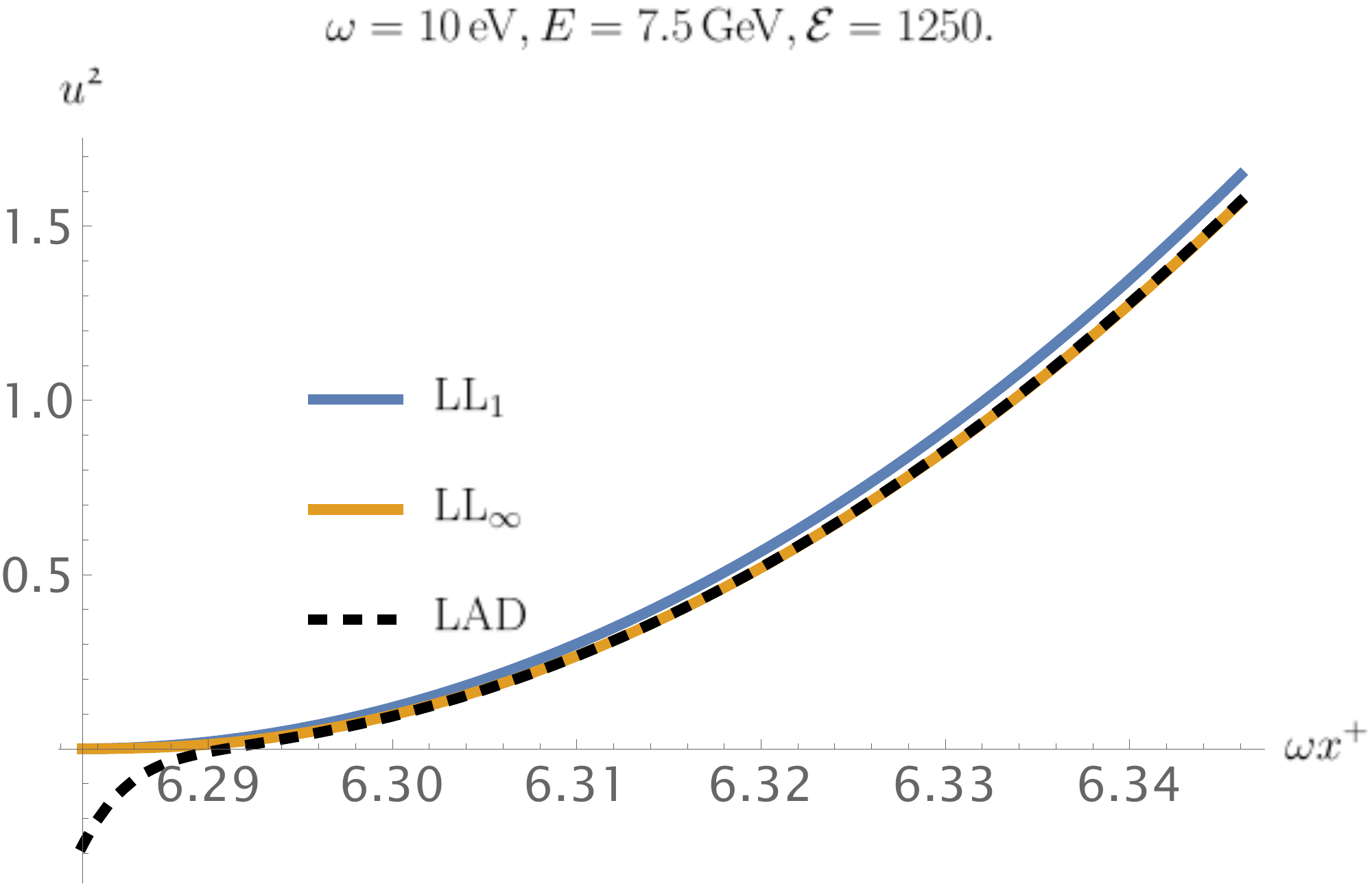}
    }
    \caption{
        Numerical solutions to $\LL1$, \LL\infty, and LAD for a particle created at the peak of the pulse~\cref{eq:pulse}.
        The latter can be extended to earlier lightfront times with the field switched off, revealing its pre-accelerating nature.
        After the field has switched on, $\LL\infty$ and LAD agree to within numerical precision.
        The dynamics of $\LL1$ differs significantly from these.
        In~(b) the disagreement in transverse components near $\omega x^\LCp = 2 \pi$ is due to numerical error, the scale of which is set by $u^\LCp \gg u^\LCperp$.
    }
    \label{fig:LLinf-LAD}
\end{figure}

We compare the solution to LAD thus obtained with that of $\LL\infty$ in \cref{fig:LLinf-LAD},
using the analytical resummants~\cref{eq:A01.Phi,eq:B01.Phi} for $\CA, \CB$.
Remarkable agreement is obtained in $u^\LCp$ despite the minimum of information used in this representation of $\LL\infty$;
the error of at most $\simeq 5\%$ near $\phi = 2 \pi$ is partly numerical, and otherwise stems from the relative error of the resummation~\cref{eq:B01.Phi} becoming significant for very large $\chi$.
(The much more apparent disagreement in $u^\LCperp$ is predominantly error propagation from $u^\LCp$.) In \cref{fig:LLinf-LAD} the low-$\chi$ regime is reached within a few percent of a cycle, well before the field changes significantly.
This justifies \emph{ex post} discarding derivative terms on this interval.
We have also solved LAD and $\LL\infty$ over one half of a cycle and seen that they still agree to within numerical precision.

As the field is only ever integrated in solving LAD, it is not a problem to explicitly turn off the field for $\phi \le 2 \pi$ and extend the LAD solution further back in time.
The pre-acceleration contained in $\iota$ is thus explicitly seen in \cref{fig:LLinf-LAD:plus},
becoming noticeable around $\Delta (\omega x^\LCp) \simeq 0.01$ before the field switches on.
Estimating $ {\Delta \tau}/{\tau_0} \approx {\Delta (\omega x^\LCp) }/{\omega \tau_0 u^\LCp(2 \pi) } \approx 3.5$,
the pre-acceleration occurs, as expected, over a few $\tau_0$ of proper time.

We mention finally that the accuracy of the approach used here can be improved further by constructing piecewise approximations to $\CA$ and $\CB$  (``stitching''), in which one uses the resummed expressions below some cutoff in $\chi$, and the convergent large-$\chi$ expansion above the cutoff.
As few as $5$ terms each for $\Phi$-Pad\'e resummation and series expansion suffice to approximate $\CA$ and $\CB$ to within a few percent over the full range of $\chi$, leading to a relative error in $u^\LCp$ of $\simeq 1\%$.
This confirms that $\LL\infty$ reproduces the physical solution of LAD, while being explicitly causal and second-order in time.

\section{Conclusions}
\label{sec:conclusion}

In this paper we have applied iterated reduction of order to the LAD equation of motion in classical elecrodynamics.
This generates an infinite sequence of possible approximations to LAD which is typically truncated at leading order, yielding the well-known Landau-Lifshitz equation.
Working for simplicity with a constant crossed field background, we have investigated the ultimate fixed point of the iteration, the equation $\LL\infty$.
(The constant field approximation can be viewed as the leading order in a derivative expansion, with the extension to next-to-leading order sketched in \cref{app:derivatives}.)

We were able to map this problem to a nonlinear initial value problem for two functions $\CA, \CB$ of the invariant $\chi$, which we have approached through (i) a direct numerical solution, which provides an `exact' benchmark,
(ii) a perturbative, weak-field analysis,
(iii) resummation of the resulting asymptotic weak-field expansion,
(iv) a strong-field expansion, which is convergent.

Regarding (ii), we can understand why the weak-field perturbative expansion is divergent through a variation of Dyson's argument~\cite{PhysRev.85.631} that the perturbation series of QED has zero radius of convergence due to the instability of the vacuum against pair production when $\alpha < 0$.
We adapt this for our case as follows: since $\alpha > 0$, an accelerating particle emits radiation in the direction of its motion, and thus RR opposes the motion.
If $\alpha$ were negative, though, particles would instead \emph{gain} momentum due to the reaction, enhancing the emission of radiation.
In such a universe particles in magnetic or Coulomb fields would spiral out, not in.
This instability cannot be connected to the physical universe, with $\alpha > 0$, by a perturbative expansion around $\alpha = 0$.

Turning to (iii), Borel-Pad\'e resummation gives a good approximation to the numerically exact solution.
Using improved resummation methods, only a single perturbative coefficient, along with data from the convergent asymptotic series (iv), is needed to yield a resummant which agrees to within a few percent for all $\chi$.
We thus have an approximate analytic expression for the fixed-point, all-orders LL equation, LL$_\infty$.
It is second order in time derivatives, and can be solved numerically to arbitrary precision.
An important qualitative feature of $\LL\infty$ is that (because the function $\CB$ is always positive) $\chi$ is always driven to $0$.
As a result large-time dynamics are governed by $\LL1$, and asymptotic quantities match those of $\LL1$.
(In particular the motion becomes aligned with the `radiation-free direction' in which there is no acceleration transverse to the motion, minimising radation losses, as also holds for LAD~\cite{Kazinski:2010ce,Kazinski:2013vga,Gonoskov2018,Ekman:2021vwg}.) Without resummation, though, it follows that the the practical, optimal, order of truncation for the series generated by iterated reduction of order is $1$.

Interestingly, though, we note that the short-time dynamics of $\LL\infty$ depends on whether it is, in the composite parameter $\chi = u^\LCp \CE$, the energy $u^\LCp$ or the intensity $\CE$ which is made large.
This is also the case in QED in strong fields, where the high-energy and high-intensity limits of observables are drastically different, scaling with powers or logarithms, respectively~\cite{Podszus:2018hnz,Ilderton:2019kqp}.

Continuing with connections to the quantum theory, a common adage is that $\LL1$ overpredicts the strength of RR compared to QED.
In this sense $\LL\infty$ gives results `closer' to QED: because the functions $\CA, \CB$ are strictly decreasing, with a maximum of $1$, $\LL\infty$ predicts less RR than $\LL1$.
This is explicitly seen in~\cref{fig:LLinf-LAD}.

Both $\LL1$ and LAD are known to be consistent with QED to leading order in $\alpha$~\cite{Krivitsky:1991vt,Higuchi:2002qc,Ilderton:2013dba}, with a very recent resummation of quantum RR in plane waves~\cite{Torgrimsson:2021wcj} recovering $\LL1$ to all orders in $\alpha$, but leading order in the pulse duration and intensity.
As this is a result for long times and in the high-\emph{intensity} limit, it is still consistent with $\LL\infty$,
which differs markedly from $\LL1$ only for short times, except in the high-\emph{energy} limit.
Making more precise, quantitative statements about the relation between $\LL\infty$, LAD, and QED requires going beyond leading order in the pulse duration or understanding the high-energy limit of strong-field QED, which remains challenging.

\begin{acknowledgments}
    The day after this manuscript was uploaded to the arXiv, Ref.~\cite{Torgrimsson:2021zob} appeared, Sec.~IV of which also treats resummation in the context of LAD.
    Where we have resummed on the level of the equation, Ref.~\cite{Torgrimsson:2021zob} has resummed on the level of the solution.
    We have checked that the approaches are consistent, yielding e.g.~the same acceleration at $x^\LCp = 0$, as the coefficients in~(60) and~(61) of Ref.~\cite{Torgrimsson:2021zob} match our~\cref{pert-coeffs}.
    We thank Greger Torgrimsson for in-depth discussions on this.

\emph{
The authors are supported by the Leverhulme Trust (RE, AI, TH), grant RPG-2019-148.
}
\end{acknowledgments}

\bibliography{LLinf}


\onecolumngrid
\appendix
\label{SECT:APPENDIX}

\section{ \texorpdfstring{$\LL\infty$}{LL∞} to leading order in derivatives }
\label{app:derivatives}

The characterisation of $\LL\infty$ as a fixed-point of reduction of order is not limited to a constant field, which serves only to limit the tensor that can appear.
In a non-constant field $\LL\infty$ must contain, among others, terms with arbitrary number of derivatives,
\begin{equation}
    \dot{u}^\mu
    =
    \sum_{\ell = 0} \CA_\ell (u \cdot \partial)^\ell f^{\mu\nu} u_\nu
    + \CB_\ell [\mathcal{P} (u \cdot \partial)^\ell f^2]^{\mu\nu} u_\nu
    + \ldots
\end{equation}
where the functions $\CA_\ell, \CB_\ell$ depend on not just the one invariant $y$, but also on, among others, $y_\ell \propto [(n \cdot \partial)^\ell f_{\mu\nu} u^\nu]^2 $.
The fixed-point condition is then, in general, an infinite tower of PDE:s in an infinite-dimensional space.

Truncating the expansion at $\ell = 0$ is justified as long as $\omega \tau_0$ remains small, where $\omega$ is a typical frequency of the field, \emph{and} the $\CA_\ell, \CB_\ell$ don't blow up.
We cannot prove this in full generality, but we can consider the next simplest case in the derivative expansion.

To first order in derivatives $\LL\infty$ must have the form
\begin{equation}
    \dot{u}_\lambda
    =
    (A_1 + z B_1) f^{\mu\nu} u_\nu
    + \frac{1}{\tau_0} (y A_2 + z B_2) (\frac{n^\mu}{n \cdot u} - u^\mu)
    + \tau_0 A_3 f^{\mu\nu,\rho} u_\nu u_\rho
\end{equation}
where the invariant $z$ is
\begin{equation}
    z = \tau_0^3 u_{\mu} f^{\mu\nu} f_{\nu\rho,\sigma} u^\rho u^\sigma
    =
    \tau_0^3 (n \cdot u)^3 (a'^\perp \cdot a''^\perp)
\end{equation}
and the coefficient functions $A_i, B_i$ depend only on $\chi$.
The initial conditions are set by $\LL1$ as $A_3(y = 0) = 1$ and a simple calculation to find $B_1(0) = -12, B_2 = 4$.

\begin{figure}[tb]
    \centering
    \includegraphics[width=0.5\linewidth]{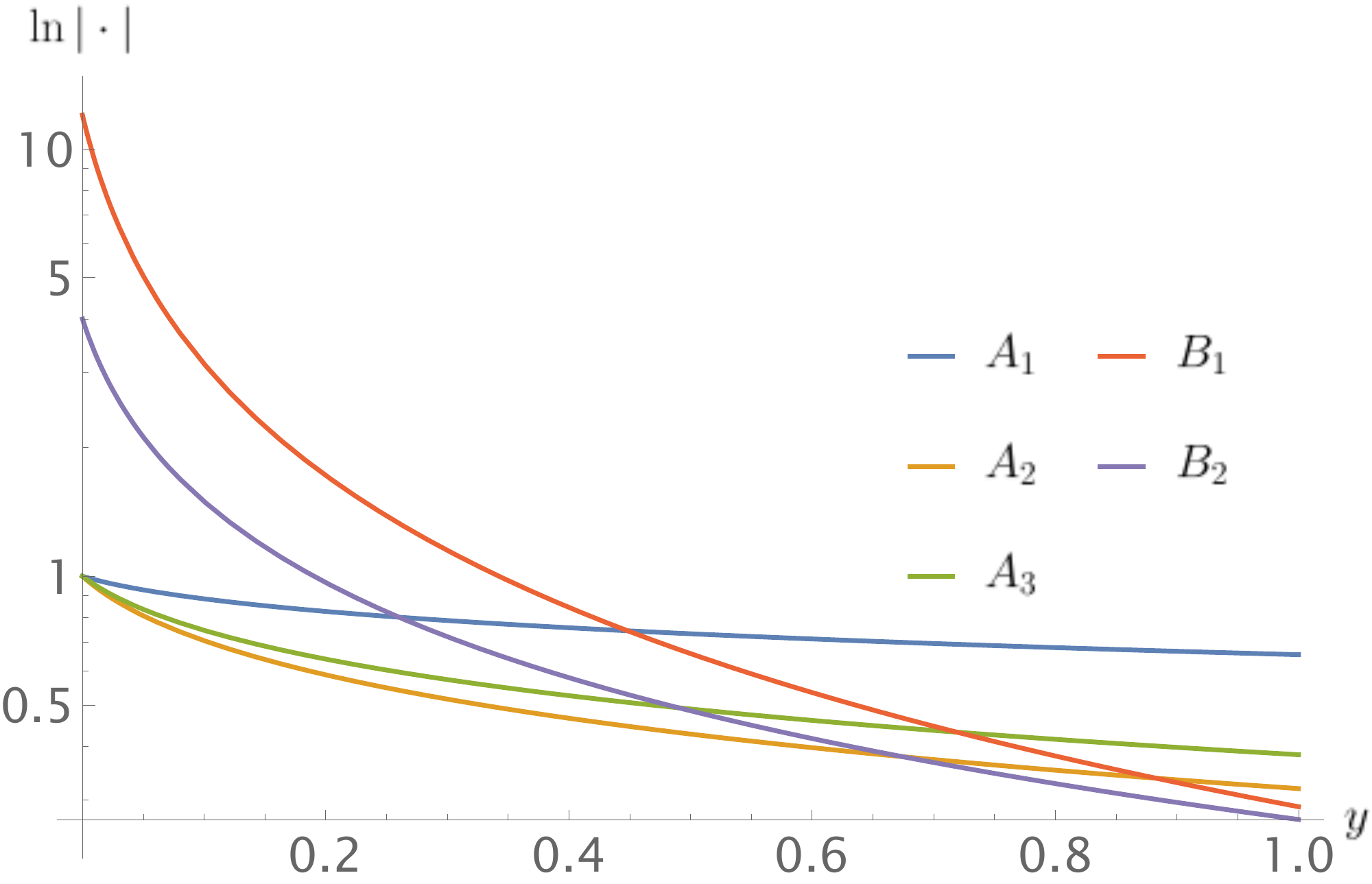}
    \caption{
        Numerical solutions for the coefficient functions $A_i$ in the leading-order derivative expansion of $\LL\infty$,~\cref{eq:derivative-ivp-eqs}--\cref{eq:derivative-ivp-ics}.
    }
    \label{fig:derivative-expansion}
\end{figure}

Applying reduction of order, identifying terms by tensor structure and order by order in $z$, the fixed-point condition now results in the initial value problem
\begin{equation}
    \label{eq:derivative-ivp-eqs}
    \begin{split}
        -2y^2 A_2 A_1'- 2y A_1 A_2 + 1
            & = A_1 \\
        -2 y^2 A_2 A_2'-2 y A_2^2 + A_1^2
            & = A_2 \\
        A_1 - 2 y^2 A_2 A_3' - 3 y A_2 A_3
            & = A_3 \\
        2(1 - y B_2) A_1' - 2y^2 A_2 B_1' - 5 y A_2 B_1 - 2 A_1 B_2
            & = B_1 \\
        2(1 - y B_2) y A_2' + 2 A_2 - 2 y^2 A_2 B_2' - 5 y A_2 B_2 + 2y A_1 B_1 + 2 A_1 A_3
            & = B_2
    \end{split}
\end{equation}
\begin{equation}
    \label{eq:derivative-ivp-ics}
    A_1(0) = A_2(0) = A_3(0) = 1 \quad B_1(0) = -12 \quad B_2(0) = 4
    \,
    .
\end{equation}
The first two equations and initial conditions are, naturally, precisely the fixed-point conditions at zeroth order in derivatives.
This system can be solved numerically, \cref{fig:derivative-expansion}, showing that all four functions fall off monotonically with $y$.
We can determine their asymptotics for $y \gg 1$ by assuming a power law falloff, resulting in
\begin{equation}
    A_1  \sim \frac{2^{1/4}}{3} y^{-1/4} \quad
    A_2  \sim \frac{2^{3/4}}{\sqrt{3}} y^{-3/4} \quad
    A_3  \sim \frac{1}{2^{3/2}} y^{-1/2} \quad
    B_1  \sim -\frac{1}{2^{3/4}} y^{-3/2} \quad
    B_2  \sim \frac{1}{3} y^{-1}
    \,
    .
\end{equation}

Furthermore, the asymptotic forms suggest that $A_3, A_4$ also have convergent series expansion around $\chi = \infty$.
As the structure of the initial-value problem is the same as without derivatives, a recursion relation similar to~\cref{eq:akbk-recur} can be found,  and
the proof in~\cref{app:proof} will apply.

\section{Proof of convergence of the series around \texorpdfstring{$\chi = \infty$}{χ = ∞} }
\label{app:proof}

\newcommand{\kdelta}{\ensuremath{\delta\big(k - {\textstyle\sum n_i}\big)}}
\newcommand{\kldelta}{\ensuremath{\delta\big(k - \ell_1 - \ell_2 - {\textstyle\sum n_i}\big)}}

We will prove that the series~\cref{eq:akbk-recur} has a finite radius of convergence.
First, assume that $k$ is large enough that we can disregard terms subleading in $k$, viz.,
\begin{equation}
    \begin{pmatrix}
        a_k \\ b_k
    \end{pmatrix}
    \approx
    -\frac{1}{3k}
    \begin{pmatrix}
        \sum \ell a_\ell b_{k - \ell} \\
        \sum \ell b_\ell b_{k - \ell}
    \end{pmatrix}
    + \mathcal{O}(1/k)
    \label{eq:ak-subleading}
\end{equation}
Let $\beta_k$ be the sequence
\begin{equation}
    \beta_k = \frac{1}{2\lambda k} \sum_{\ell=1}^{k-1} \ell \beta_\ell \beta_{k - \ell}
    =
    \frac{1}{4\lambda} \sum_{\ell=1}^{k-1} \beta_\ell \beta_{k - \ell}
    \quad
    k > 1
    \quad
    \,
    .
    \label{eq:btk-def}
\end{equation}
If $|\beta_\ell| > |b_\ell|, |a_\ell|$ for $\ell < k$, the derivative term is bounded by $\beta_k$, so $a_k,b_k$ differ by at most $\mathcal{O}(1/k)$.
Thus $\beta_k$ will bound $|a_k|, |b_k|$ as long as it does so until $k$ becomes sufficiently large.
We will now prove that through induction on $k$ that
\begin{equation}
    \beta_k \le \lambda \frac{\beta_1^k}{\lambda^k}
    \label{eq:induc-hyp}
    \,
    ;
\end{equation}
the base case $k=1$ is obviously true.
To reduce clutter let's introduce the shorthand
\begin{equation}
    [n_1, n_2, \cdots, n_N] := \prod_{i=1}^N \beta_{n_i}
\end{equation}
First note that $\beta_2 = \beta_1^2/(4\lambda)$; then for $k > 2$ write
\begin{align}
    \beta_k
    & = \frac{\beta_1}{2\lambda} \beta_{k-1}
        + \frac{1}{4\lambda} \sum_{n_i \ge 2} [n_1, n_2] \kdelta
    \\
    & = \frac{\beta_1}{2\lambda} \beta_{k-1}
        + \frac{1}{(4\lambda)^2} \sum_{n_1 \ge 2} [\ell_1, \ell_2, n_1] \kldelta
\end{align}
``splitting'' either of the summation indices using a the definition~\cref{eq:btk-def}.
Since $n_1 \ge 2$ the maximally indexed factors that appear in the triple product are $\beta_{k-3}$ and $\beta_{k-4}$, according to the minimal sums of $\ell_i$.
The former has combinatorical weight $1$ and the latter $2$.
Thus,
\begin{equation}
    \beta_k
    = \frac{\beta_1}{2\lambda} \beta_{k-1}
        + \frac{\beta_1^2}{8\lambda^2} \beta_{k-2}
        + \frac{\beta_1^3}{16\lambda^3} \beta_{k-3}
        + \frac{1}{(4\lambda)^3} \sum_{n_i \ge 2} [n_1, n_2, n_3] \kdelta
\end{equation}
and the triple product can be split into a quadruple product with maximal indices $k-4$ and $k-5$.
The combinatorical weights are $1 \times 2$ and $2 \times 2$, respectively.
In general, splitting an index in a product with $N$ indices (which will have a factor $(4\lambda)^{-N+1}$ from previous splits, will work out like
\begin{align}
    \sum_{n_i\ge 2} [n_1, \cdots, n_N] \kdelta
    & =
    \frac{1}{4\lambda} \sum_{n_i\ge 2} [\ell_1, \ell_2, n_1, \cdots, n_{N-1}] \kldelta
    \\
    &= \frac{N-1}{4\lambda} \beta_2^{N-2} (
        \beta_1^2 \beta_{k - 2N + 2} + 2 \beta_1 \beta_2 \beta_{k - 2N - 1}
        ) + \frac{1}{4\lambda} \sum_{n_i} [n_1, \ldots, n_{N+1}] \kdelta
    \\
    & \le
    \frac{\beta_1^{2N-2}}{2^{N-1}\lambda^{N-1}} \beta_{k-2N+2} + \frac{\beta_1^{2N-1}}{2^{N-2}\lambda^{ N }} \beta_{k-2N-1}
     + \frac{1}{4\lambda} \sum_{n_i} [n_1, \ldots, n_{N+1}] \kdelta
\end{align}
Repeating and using the induction hypothesis~\cref{eq:induc-hyp} we can bound $\beta_k$ by a geometric series:
\begin{equation}
    \label{eq:geo-series}
    \beta_k \le \sum_{\ell=1} \frac{\beta_1^\ell}{(2\lambda)^\ell} \beta_{k-\ell}
    \le \lambda \frac{\beta_1^k}{\lambda^k} \sum_{\ell=1} \frac{1}{2^\ell}
    = \beta_1 \frac{\beta_1^k}{\lambda^k}
    \,
    .
\end{equation}
\newcommand{\betat}{\tilde{\beta}}
To get a bound for our $b_k$:s we should take $\beta_1 = |b_1| = \frac{9}{2}$ and $\lambda \le 3/2$ which implies $\rho \ge \frac{3}{2 |b_1|} = 1/3$.
As evidenced in \cref{fig:partial-sums}, the actual radius of convergence is larger.
This is because $a_\ell, b_\ell$ are more or less alternating.
To take this into account, we can consider the analogous sequence $\betat_k$ where we set $\betat_2 = -\betat_1^2/4\lambda$.
Now realise that every time we pick an $\ell_i = 2$, we get a minus sign from this, so the geometric series~\cref{eq:geo-series} becomes alternating, and we can actually bound
\begin{equation}
    |\betat_k| \le \lambda \Big( \frac{|\betat_1|}{2\lambda} \Big)^k
    \,
    .
\end{equation}
Thus, our sequences $a_k, b_k$ are bounded by
\begin{equation}
    \Big| \frac{a_k}{a_1} \Big|,
    \Big| \frac{b_k}{b_1} \Big|
    \le r^k
\end{equation}
as long as $r$ is large enough to both satisfy $r > \frac{1}{2} \max ( |a_1|, |b_1| )$ and provide a bound even for $1 \centernot\ll k$.
We can estimate such an $r$ by a linear fit to a few hundred elements, or by graphical estimation; see again~\cref{fig:akbk-bounded}:
if $r$ is such that all the points are below its trendline, the radius of convergence in $(\frac{3\tau_0}{\sqrt{2}} \chi)^{-1/2}$ is at least $1/r$.
We see that $r = \frac{3}{2}$ fails to give a bound for the $a_k, b_k$ around $k \approx 50$, but $r \approx 1.7$ -- found by fitting -- provides the bound~\cref{eq:radius-of-convergence}, $m\tau_0 \chi \gtrsim 1.37$.

\section{Numerical scheme for LAD}
\label{app:numerics}

The runaway solutions can be removed by writing LAD in an integro-differential form; that presented in Ref.~\cite{PhysRevE.88.033203} is suitable.
It is obtained by rearranging LAD into
\begin{equation}
    f^{\mu\rho} u_\rho + \tau_0 u^\mu \dot{u}^\rho \dot{u}_\rho
    =
    -\tau_0 e^{\tau/\tau_0} \frac{\ud}{\ud \tau} \left( e^{-\tau/\tau_0} \dot{u}^\mu \right)
    \,
\end{equation}
tensoring by $u^\nu$ and anti-symmetrising to kill the term quadratic in $\dot{u}^\mu$, viz.,
\begin{equation}
    -\frac{e^{-\tau/\tau_0}}{\tau_0} ( f^{\mu\rho} u_\rho u^\nu - f^{\nu\rho}u_\rho u^\nu )
    =
    u^{[\nu} \frac{\ud}{\ud \tau} \left( e^{-\tau/\tau_0} \dot{u}^{\mu]} \right)
    =
    \frac{\ud}{\ud \tau} \left( e^{-\tau/\tau_0} u^{[\nu} \dot{u}^{\mu]} \right)
    \,
    .
\end{equation}
The equation is now solved by formal integration ($\ud x^\LCp = x^\LCp \, \ud \tau$) and dotting with $u^\nu(x^\LCp)$,
\begin{equation}
    \label{eq:alcaine}
    \frac{\ud u^\mu}{\ud x^\LCp}
    =
    \frac{e^{\tau/\tau_0}}{\tau_0} \frac{u_{\nu}}{u^\LCp}  \int_{x^\LCp}^\infty
    \frac{\ud y \, e^{-\tau(y)/\tau_0}}{u^\LCp(y)} [ f^{\mu\rho} u_\rho u^\nu - f^{\nu\rho}u_\rho u^\mu ](y)
    \,
    ,
\end{equation}
free of runaways as long as $f^{\mu\nu}$ vanishes sufficently quickly.

\newcommand{\ttau}{\tilde{\tau}}
Our discretisation scheme is a simple finite difference scheme.
Letting $\ttau = \tau/\tau_0$, and subscript $k$ indicating evalua\-tion at $\phi_k$, we proceed according to
\begin{equation}
    u^\mu_{k+1} - u^\mu_k
    =
    \Delta\phi \frac{1}{\tau_0 \omega} \frac{u^{\nu}_k}{u^\LCp_k}  \int_{\phi}^\infty \ud \phi' \,
    \frac{e^{(\ttau(\phi)-\ttau(\phi')}}{u^\LCp(\phi')} [ \tilde{f}^{\mu\rho} u_\rho u_\nu - \tilde{f}^{\nu\rho}u_\rho u^\mu ](\phi')
    \qquad
    \tilde{f}^{\mu\nu} = n^{[\mu} \frac{\ud a^{\nu]}}{\ud \phi}
\end{equation}
where we evaluate the integral by the trapezoid method,
\begin{align}
    I_k = \int_{\phi_k}^\infty e^{\ttau_k - \ttau(y)} \cdots \, \ud y
    & =
    \int_{\phi_k}^{\phi_{k-1}} e^{\ttau_k - \ttau(y)} \cdots \, \ud y
    + e^{\ttau_k - \ttau_{k-1}} \int_{\phi_{k-1}}^\infty  e^{\ttau_{k-1} - \ttau(y)} \cdots \, \ud y
    \\
    & \approx
    \frac{\Delta \phi}{2}\left( f(\phi_k, u_k )  + e^{\ttau_k - \ttau_{k-1}} f(\phi_{k-1}, u_{k-1} ) \right)
    + e^{\ttau_k - \ttau_{k-1}} I_{k-1}
    \Big)
    \,
    .
\end{align}
The trapezoid method is also used to evaluate the proper time step,
\begin{equation}
    \ttau_k - \ttau_{k-1} = \frac{\tau_k - \tau_{k-1}}{\tau_0}
    =
    -\frac{\Delta\phi}{2\omega \tau_0} \left( \frac{1}{u^\LCp_k} + \frac{1}{u^\LCp_{k-1}} \right)
    \,
    ;
\end{equation}
while $I_0 = \int_{\phi_0}^\infty \ud y \, \cdot $ is evaluated using a solution of $\LL1$, since by assumption all later times are in the low-$\chi$ regime where $\LL1$ and LAD agree;
this initialisation decays over a very short phase interval, as $u^\LCp$ is small near $\phi_0$, and the solution is not sensitive to it.
While the equation is written in covariant form, we treat only $u^\LCp, u^\LCperp$ as independent, fixing the final component through the mass-shell condition $u^\LCm = \frac{1 + (u^\LCperp)^2}{u^\LCp}$.

\end{document}